\documentclass[%
 reprint,
 superscriptaddress,
 amsmath,amssymb,
pnas,
]{revtex4-1}

\usepackage{graphicx}
\usepackage{dcolumn}
\usepackage{bm}
\usepackage{makecell}
\usepackage{epsfig}
\usepackage{subcaption}

\usepackage[mathlines]{lineno}

\begin{document}


\title{Sediment load determines the shape of rivers}
\author{Predrag Popovi\'{c}}
 \altaffiliation{All correspondence should be directed to Predrag Popovi\'{c}}
  \email{arpedjo@gmail.com}
  \affiliation{%
 Universit\'e de Paris, Institut de Physique du Globe de Paris, 1 rue Jussieu, CNRS, F-75005 Paris, France\\
}%
\author{Olivier Devauchelle}%
\affiliation{%
Universit\'e de Paris, Institut de Physique du Globe de Paris, 1 rue Jussieu, CNRS, F-75005 Paris, France\\
}%
\author{Ana\"{i}s Abramian}
\affiliation{%
Sorbonne Universit\'e, CNRS - UMR 7190, Institut Jean Le Rond d'Alembert, F-75005 Paris, France\\
}%
\author{Eric Lajeunesse}
\affiliation{%
Universit\'e de Paris, Institut de Physique du Globe de Paris, 1 rue Jussieu, CNRS, F-75005 Paris, France\\
}%




\date{\today}

\begin{abstract}

Understanding how rivers adjust to the sediment load they carry is critical to predicting the evolution of landscapes. Presently, however, no physically based model reliably captures the dependence of basic river properties, such as its shape or slope, on the discharge of sediment, even in the simple case of laboratory rivers. Here, we show how the balance between fluid stress and gravity acting on the sediment grains, along with cross-stream diffusion of sediment, determines the shape and sediment flux profile of laminar laboratory rivers which carry sediment as bedload. Using this model, which reliably reproduces the experiments without any tuning, we confirm the hypothesis, originally proposed by Parker \cite{parker1978self}, that rivers are restricted to exist close to the threshold of sediment motion (within about $20\%$). This limit is set by the fluid-sediment interaction and is independent of the water and sediment load carried by the river. Thus, as the total sediment discharge increases, the intensity of sediment flux (sediment discharge per unit width) in a river saturates, and the river can only transport more sediment by widening. In this large discharge regime, the cross-stream diffusion of momentum in the flow permits sediment transport. Conversely, in the weak transport regime, the transported sediment concentrates around the river center without significantly altering the river shape. If this theory holds for natural rivers, the aspect ratio of a river could become a proxy for sediment discharge --- a quantity notoriously difficult to measure in the field.

\end{abstract}

\maketitle

Flowing from mountains to oceans, rivers traverse immense distances across the land, eroding, transporting, and depositing sediment along the way, thereby shaping much of the landscape we see on Earth \cite{schumm1979geomorphic,seminara2010fluvial,metivier2012alluvial,phillips2016self}. However, a precise understanding of how rivers adjust their shape to the amounts of sediment and water they transport is lacking. This is partly due to the difficulty of collecting sediment flux measurements in the field, and partly due to the complicated coupling between the flow and the sediment bed.

In rivers that carry a small amount of sediment, sediment grains are typically close to their threshold of motion --- below this threshold, any sediment carried by a river would be deposited, building the river bed until it eventually reaches the threshold, while, above the threshold, uncompensated erosion of the bed would quickly bring the river back to the threshold \cite{parker1976cause}. For this reason, early theories were formulated for inert rivers (rivers that do not transport sediment) and assumed that such rivers construct their own bed so that the grains on the bed surface are exactly at the threshold of motion \cite{glover1951stable,henderson1961stability,parker2007physical,savenije2003width}. Under this assumption, they showed that the shape of the river channel is independent of its water discharge, which can only affect the size of the river. This threshold theory accounts for the observation that the width of rivers increases as the square root of their discharge, an empirical correlation known as Lacey's law \cite{savenije2003width,metivier2017laboratory}.

In active rivers (those that transport sediment), sediment transport is driven only by a small departure of the shear stress from its threshold value \cite{phillips2016self,parker2007physical}. The minuteness of this departure makes the study of active rivers challenging. It means that, to find the sediment flux, one needs to measure or calculate the stress with high precision --- simple order-of-magnitude estimates are not sufficient \cite{phillips2019bankfull}. This is a daunting task, since the stress sensitively depends on the river shape, which, in turn, adjusts to the stress distribution. 

Parker \cite{parker1978self} first addressed the question of active rivers with a model in which a turbulent river splits into inert banks and a flat, active bottom. He found that the cross-stream diffusion of momentum, which distributes stress from faster flowing regions to slower ones, is essential to enable sediment transport in a stable river channel. His model qualitatively agreed with real rivers --- he found that the stress on the river bed is at most about $20\%$ above critical, which limits the intensity of sediment transport. It is, however, unclear why a river should sharply split into inert banks and a flat bottom, as required by Parker's model. Moreover, it is unclear how a river transitions from an inert, threshold channel to a singular configuration of Parker as its sediment discharge increases. 

Since field measurements are difficult, a good place to test our understanding of rivers is the laboratory \cite{malverti2008small}. However, even laboratory investigations have been a challenge in themselves \cite{ikeda1981self,schumm1987experimental,federici2003dynamics} --- stable single-thread rivers were only recently produced in a laboratory setting \cite{seizilles2013width,abramian2018self,abramian2020laboratory}. Nevertheless, these experiments have been enlightening --- by focusing on straight, laminar, stationary rivers, they presented strong support for the threshold hypothesis of inert rivers. So far, however, they have not been compared to Parker's theory for active rivers. 

Another key insight that arose from experiments is that the grains that are carried as bedload (i.e. that are dragged along the river bed) \cite{bagnold1973nature} diffuse laterally by randomly hitting other grains on the bed surface along their way \cite{seizilles2014cross,abramian2019boltzmann}. In analogy with a gas placed in a gravitational field, the balance between gravity and diffusion distributes the transported grains over the bed so that the concentration of moving grains exponentially falls off with increasing elevation above the channel centerline \cite{abramian2019boltzmann}. This Boltzmann distribution of the moving grains relates the sediment flux to the shape of the river. The role of sediment diffusion was recognized early in rivers that transport their sediment in suspension \cite{parker1978selfpart1}, but these experiments have shown that this mechanism also applies to bedload transport.

In this paper, we use the experiments of Abramian et al. \cite{abramian2020laboratory} (section \ref{sec:experiment}) to understand what sets the channel shape of active rivers. In our theory, the above mechanisms combine to shape the river --- the shape of the channel determines the stress, the stress determines the sediment flux, while the Boltzmann distribution relates the sediment flux back to the shape (section \ref{sec:mechanisms}). In equilibrium, these mechanisms are all coupled together, and their simultaneous coexistence determines a unique river channel for given discharges of water and sediment (assuming the channel is straight and single-thread). Therefore, the problem can be solved self-consistently, at least in principle. However, this problem is difficult since the fluid stress anywhere on the bed depends on the entire shape of the river. 

We bypass this issue by simplifying the equation for the fluid stress, assuming that the aspect ratio of a river (ratio of width to depth) is large (section \ref{sec:bvp}). We then formulate a model for the steady-state shape of a straight, laminar river with bedload transport by using this minimal representation for the stress, and including the Boltzmann distribution for the moving grain density. This model takes the form of a second-order boundary value problem (BVP) which can be analyzed numerically (section \ref{sec:dependence_on_discharges}) and analytically (section \ref{sec:phase_space}). We note that this is a well defined problem only for a river in equilibrium (steady-state), so that it does not answer how the river reaches this equilibrium. 

In the limit of large water and sediment discharge, the river in our model splits into inert banks and a flat active bottom, exactly as prescribed in Parker's \cite{parker1978self} model. We, thus, show how Parker's \cite{parker1978self} model arises as a limit of our theory. We call this limit the ``Parker regime'' (section \ref{sec:Parker}), and we define a condition for reaching it. Like Parker \cite{parker1978self}, we find that laminar rivers cannot exist far from the threshold of sediment motion and, thus, cannot accommodate a sediment flux (discharge per unit width) larger than a maximum. We find that this maximum depends only on the friction coefficient of the sediment, $\mu_t$. Since the sediment flux is bounded, a river in the Parker regime has to widen to accommodate a larger sediment discharge. Moreover, we find that momentum diffusion in the flow plays a key role in sediment transport. We compare our results with laboratory experiments and find good agreement without any tuning. In this way, for the first time, we provide support for Parker's hypothesis in a controlled setting. However, we also discover another qualitatively different regime, which applies to rivers with large water but small sediment discharge. In this ``weak transport regime'' (section \ref{sec:weak_transport}), sediment transport relies on the diffusion of sediment, and a higher load is accommodated by increasing the sediment flux without altering the shape of the river which carries it. 

In both theory and experiments, we find that the aspect ratio of a river strongly depends on its sediment discharge. This suggests that, in the field, the shape of the river could be used as a proxy for its sediment load. To verify this, however, our theory would have to be adapted for turbulent flows --- a task we leave for the future. 

We also leave the mathematical details, tables for experimental runs, and other results that are not necessary to understand the main points of the paper to the Supplementary Information (SI).

\section{Experiments}\label{sec:experiment}

\begin{figure}
\begin{subfigure}{1.\columnwidth}
\centering
\includegraphics[width=1.\linewidth]{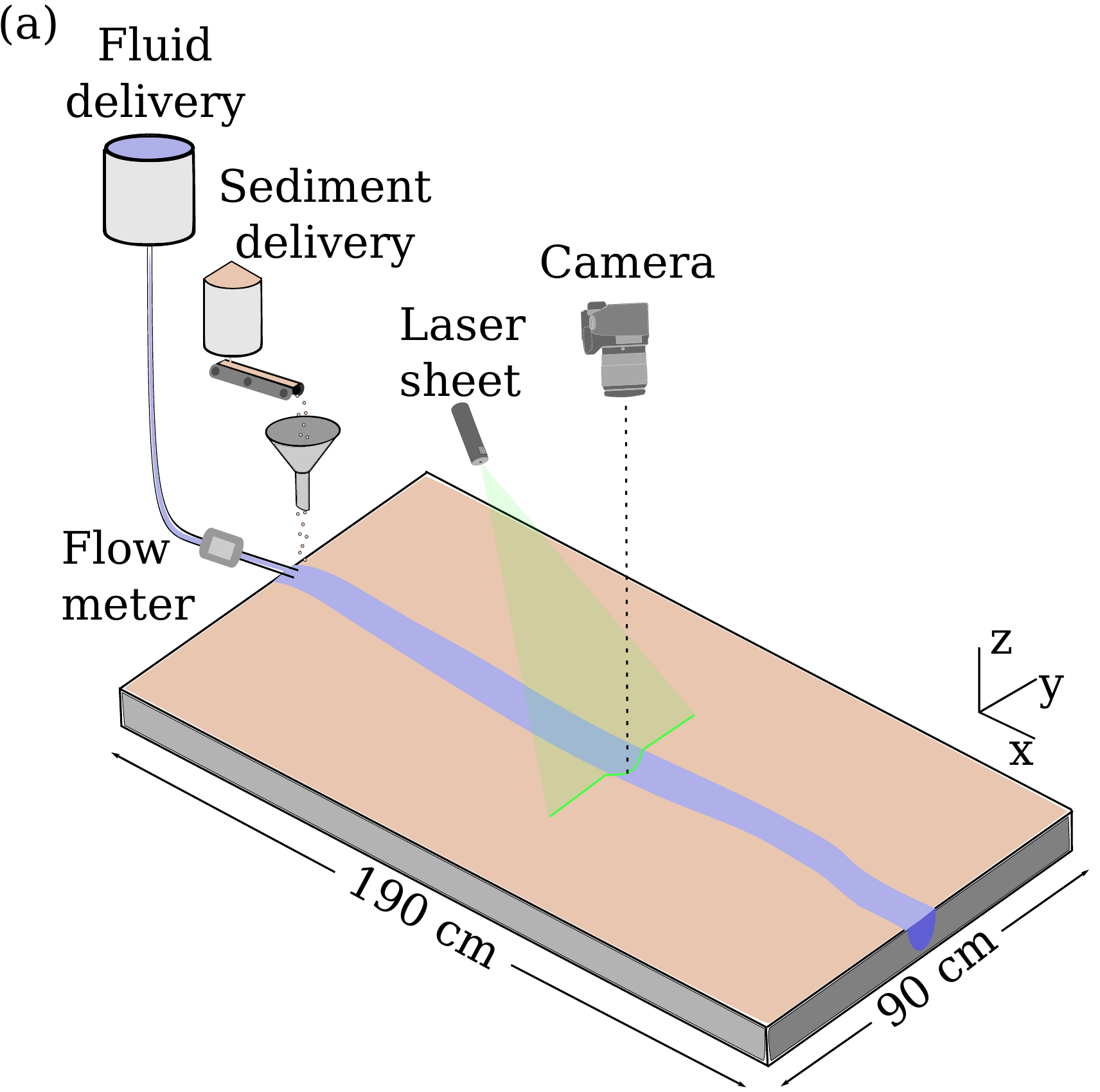}
\end{subfigure}
\newline
\begin{subfigure}{1.\columnwidth}
\centering
\includegraphics[width=1.\linewidth]{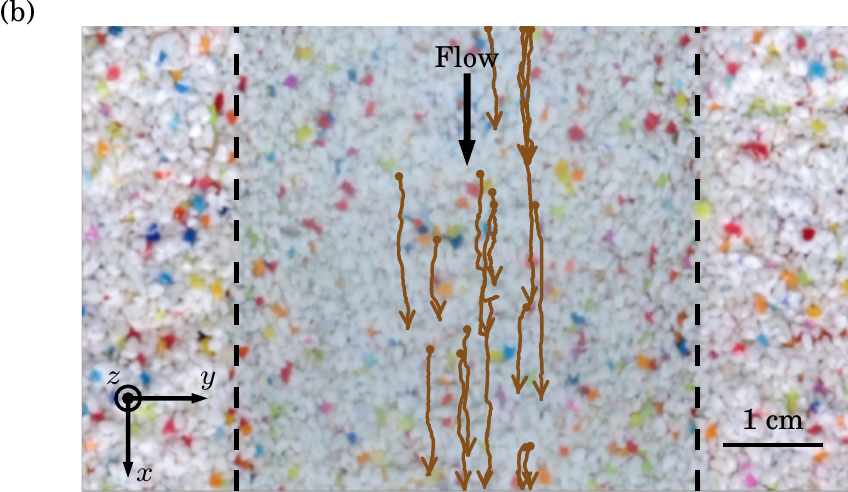}
\end{subfigure}
\caption{ (a) Experimental setup of Abramian et al. \cite{abramian2020laboratory}. (b) Photograph of the sediment bed taken with the overhead camera. Brown lines represent trajectories of tracked grains. }
\label{fig:1}
\end{figure}

In this section, we briefly describe the experiments of Abramian et al. \cite{abramian2020laboratory}, which inspired the present theory. A schematic and a photograph of the experiment are shown in Fig. \ref{fig:1}, and experimental parameters are summarized in Table S1 of the SI.

The setup consisted of an inclined tank, $190 \text{ cm} \times 90 \text{ cm} \times 10 \text{ cm}$ in size, filled with plastic sediment made up of grains of diameter $d_s = 0.83 \pm 0.2 \text{ mm}$ and density $\rho_s = 1490 \text{ kg m}^{-3}$. At the inlet, a mixture of water and glycerol was pumped into the tank, at a discharge $Q_w \approx 1 \text{ l min}^{-1}$, which was kept as constant as possible during all experimental runs. The density and viscosity of the fluid were $\rho_f = 1160 \pm 5 \text{ kg m}^{-3} $ and $\nu = 10^{-5} \text{ m}^2\text{ s}^{-1}$. The high fluid viscosity, achieved by adding glycerol to the mixture, kept the fluid flow laminar (the Reynolds number remained below about 10 in all experiments). The fluid found its way to the outlet at the opposite end of the tank, meanwhile carving its own channel through the sediment. Additionally, dry sediment was injected into the system at a prescribed rate, $Q_s$. Abramian et al. \cite{abramian2020laboratory} performed 5 experimental runs in which they varied the sediment discharge between $0$ and $60 \text{ grains s}^{-1}$. 

\begin{figure*}
\centering
\includegraphics[width=1.\textwidth]{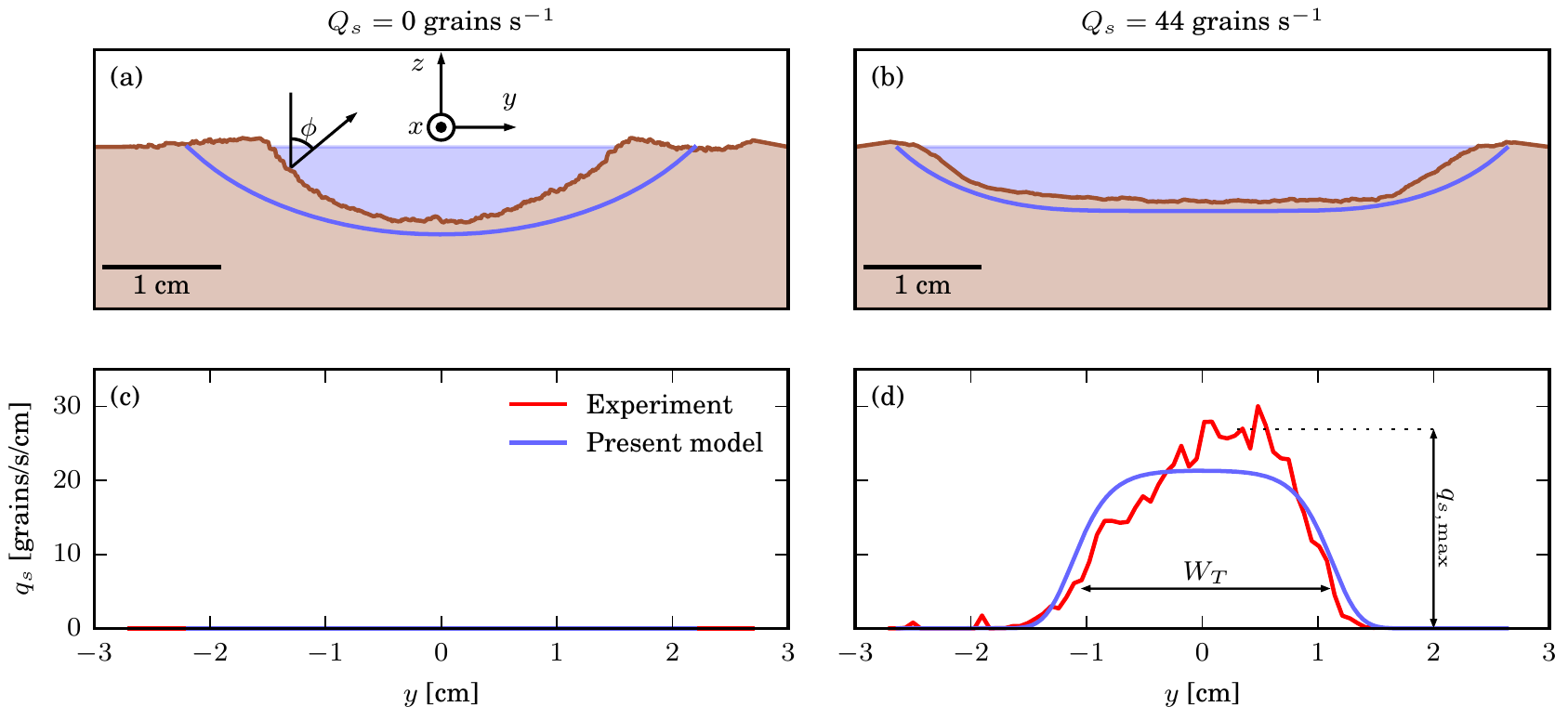}
\caption{(a) and (b) River cross-sections from the experiments of Abramian et al. \cite{abramian2020laboratory} (brown line) and the present model (blue lines). Aspect ratio is preserved. (c) and (d) Corresponding sediment flux profiles, $q_s(y)$, for the experiments (red lines) and our model (blue lines). Panels on the left ((a) and (c)) correspond to an inert river (no sediment discharge, $Q_s = 0$), while the right panels ((b) and (d)) correspond to an active one (sediment discharge $Q_s \approx 44 \text{ grains s}^{-1}$). The transport width, $W_T = Q_s / \langle q_s \rangle$, with $\langle q_s \rangle$ given by Eq. \ref{mean_qs_def}, as well as the maximum sediment flux, $q_{s, \mathrm{max}}$, are marked with arrows in panel (d). The downstream slope, $S$, could not be measured accurately, but it is approximately $S\approx 0.005$ for the inert river and $S\approx0.01$ for the active one.}
\label{fig:2}
\end{figure*}

\begin{figure*}[!th]
\includegraphics[width=0.66\textwidth]{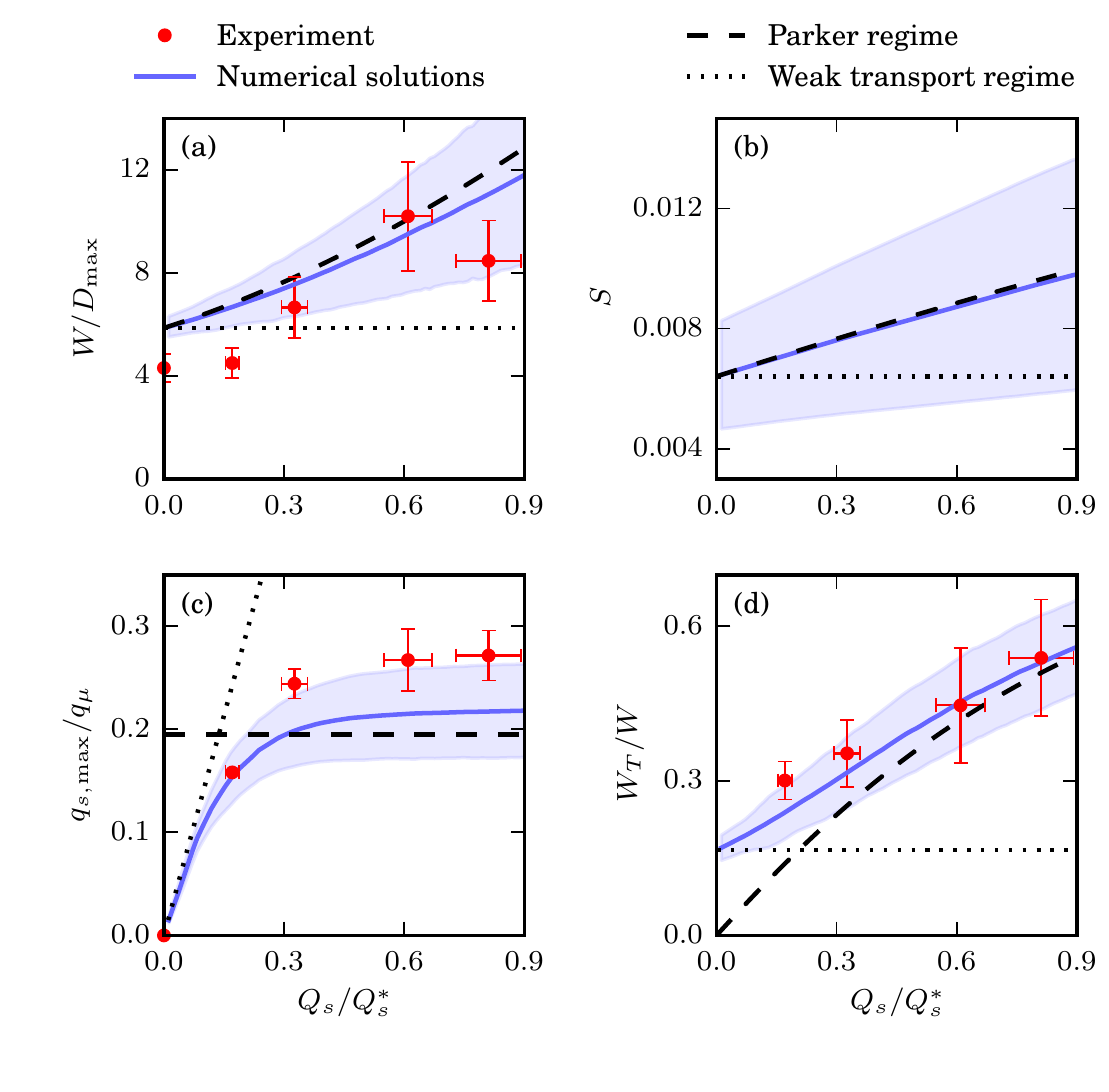}
\caption{ River properties as a function of the sediment discharge, $Q_s$, normalized by the characteristic discharge, $Q_s^* \approx 74 \text{ grains s}^{-1}$, given by Eq. \ref{Parker_Qs_characteristic}. Red dots represent the experiments (error bars estimated in SI section S1). Blue lines represent the numerical solutions to Eq. \ref{BVP} using the experimental parameters (see Table S1 of the SI). Light blue shading corresponds to the uncertainty in the parameter estimates (Table S1 of the SI). The numerical solutions transition from the weak transport regime (black dotted line) to the Parker regime (black dashed line) when $Q_s \sim Q_{s,\text{t}} \approx 8.6 \text{ grains s}^{-1}$ (Eq. \ref{Qs_transition}). (a) River aspect aspect ratio, $W/D_\text{max}$. The weak transport regime assumes a fixed bed shape so the aspect ratio is constant. (b) Downstream slope, $S$. The slope is too small for direct measurement. As in panel (a), the fixed bed shape in the weak transport regime leads to a constant slope, while the Parker regime follows from Eq. \ref{Parker_L}. (c) Normalized maximum sediment flux, $q_{s, \text{max}} / q_\mu$, where $q_\mu$ is the prefactor of the sediment transport law (Eq. \ref{qs_mu}). The weak transport regime corresponds to Eq. \ref{Gauss_qsmax} while the Parker regime corresponds to Eq. \ref{Parker_qs}. (d) Transport width, $W_T \equiv Q_s/ \langle q_s \rangle$, normalized by the total width, $W$. The weak transport regime corresponds to Eq. \ref{Gauss_WT}, while the Parker regime follows from Eqs. \ref{Parker_qs}, \ref{Parker_WT}, and \ref{Parker_L}.  }
 \label{fig:3}
 \end{figure*}

A typical river forms as follows. First, the experiment goes through a transient during which the fluid erodes more sediment than is injected at the inlet. At this stage, a single channel of width $W\sim 5\text{ cm}$ quickly forms, whose downstream slope, $S$, slowly changes over time until it reaches steady-state at $S \sim 0.01$. The duration of this transient, $T$, roughly corresponds to the time to build a sediment channel of constant slope, $S$, and width, $W$, over the entire length of the tank, $\mathcal{L} \sim 2\text{ m}$, by exchanging sediment at a rate $Q_s \sim 100 \text{ grains s}^{-1}$ with the bed. A simple scaling analysis yields $T \sim \mathcal{L} ^2 W S / (d_s^3Q_s) \sim 5 \text{ h}$, consistent with typical transients in the experiments. The exact duration of the transient depends on the initial setup of the experiment and can be shortened by, for example, setting the initial inclination of the tank close to the steady-state slope of the river. After reaching steady-state, the river transports as much sediment along its bed as is delivered by the sediment feeder. Sediment travels as bedload --- grains roll, slip, and bounce on the sediment bed. The river channel typically appears to be roughly straight with only minor sinuosity, and, once formed, it does not move significantly. Moreover, the steady-state river is insensitive to the initial setup of the experiment --- it selects its own width, $W$, depth, $D_\text{max}$, and downstream slope, $S$, regardless of the initial conditions. Beyond a certain value of sediment discharge (about $Q_s \approx 90 \text{ grains s}^{-1}$), the channel destabilizes into intertwined threads that form a braided river. The range of $Q_s$ explored in these experiments covered the entire range of sediment discharge for which a stable single-thread river can form. 

To characterize the shape of these experimental rivers, Abramian et al. \cite{abramian2020laboratory} measured the sediment bed elevation along a cross-section with a laser sheet. They constantly monitored the river using an overhead camera, and tracked the trajectories of moving colored grains, which allowed them to measure the profile of sediment flux, $q_s$, across the river (to avoid possible confusion, we emphasize here that the sediment discharge, $Q_s$, is the integral of the sediment flux, $q_s$, over the cross-section of the river). We show two rivers and their sediment flux profiles in Fig. \ref{fig:2}; profiles for the other runs are shown in Fig. S1 and their properties are summarized in Table S2 of the SI. Most sediment concentrates near the channel center over a well-defined bed section of width $W_T$. We define this transport width, $W_T$, as the width that relates the sediment discharge and the mean sediment flux, $Q_s = W_T \langle q_s \rangle $. To make $W_T$ a robust quantity resistant to experimental noise, we define $\langle q_s \rangle$ to be the average sediment flux over a probability density function $q_s/Q_s$, so that
\begin{linenomath*}
\begin{equation} \label{mean_qs_def}
\langle q_s \rangle  \equiv \frac{1}{Q_s} \int_{-W/2}^{W/2} q_s^2(y)\text{d}y \text{  . } 
\end{equation}
\end{linenomath*}  
Figure \ref{fig:3} and Fig. S2 of the SI illustrate how the characteristics of laboratory rivers change as the sediment discharge, $Q_s$, increases: the rivers become wider, shallower, steeper, and transport sediment more intensely. 

\section{The mechanisms that shape a river}\label{sec:mechanisms}

Keeping in mind the rivers of Abramian et al. \cite{abramian2020laboratory}, the goal of the present paper is to understand how an active laminar river adapts its own depth and sediment flux profiles, $D(y)$ and $q_s(y)$, to the fluid and sediment discharges, $Q_w$ and $Q_s$, it carries. In this section, we will start by reviewing the equations which govern the flow and the transport of sediment in such a river. Throughout the paper, $x$ will represent the downstream, $y$ the cross-stream, and $z$ the vertical coordinate, measured with respect to the surface of the river (Fig. \ref{fig:2}a). We restrict our attention to a straight river that is uniform in the $x$-direction. Accordingly, we only need to consider its cross-section in the $(y,z)$ plane.

\subsection*{Stokes flow}

In a straight river, the flow is forced by gravity that pushes the fluid down a slope, $S$. This slope is usually very small (for the experiments of Abramian et al. \cite{abramian2020laboratory}, $S \sim 0.01$). The laminar flow in such a river obeys the Stokes equation
\begin{linenomath*}
\begin{equation} \label{Stokes_law}
\nu \Delta u = - g S \text{  , }
\end{equation}
\end{linenomath*}
where $u$ is the downstream component of the velocity, $g = 9.81 \text{ m s}^{-2}$ is the gravitational acceleration, $S$ is the slope in the downstream ($x$) direction, and $\Delta \equiv \frac{\partial^2}{\partial y^2} +\frac{\partial^2}{\partial z^2}$ is the Laplacian operator in the $(y,z)$ plane. The boundary conditions are that the velocity vanishes on the bed ($u = 0$ when $z = -D$) and that there is no shear stress on the free surface ($\partial u / \partial z = 0$ when $z = 0$).

The term $g S$ in Eq. \ref{Stokes_law} is the force driving the fluid flow. In the experiments, the slope is not prescribed a priori. Instead, the river selects it while forming its own bed. It depends on the river's discharges and we cannot prescribe it arbitrarily. Importantly, the Stokes flow is scale-invariant --- the flow in two channels of a different size but the same shape looks the same, and one can find one from the other by simple rescaling of lengths and velocity.

If we can find the velocity in the channel by using Eq. \ref{Stokes_law}, we can also get the stress, $\tau$, shearing the bed surface. This stress is proportional to the gradient of $u$ in the direction normal to the bed surface, with the dynamic viscosity, $\rho_f \nu$, acting as a constant of proportionality. To get an idea of how the stress depends on the channel shape, we integrate Stokes law, Eq. \ref{Stokes_law}, along the vertical direction, and find an equation for $\tau$:
\begin{linenomath*}
\begin{align} \label{stress}
&\tau = \left( \rho_f g S D + \rho_f \nu (\bar{u} D)'' \right) \cos \phi \text{  , }\\
&\bar{u} \equiv \frac{1}{D}\int_{-D}^0  u \text{d} z \text{  , }
\end{align}
\end{linenomath*}
where primes denote $y$-derivatives, $\bar{u}$ is the vertically averaged flow velocity, and $\phi$ is the angle between the vector normal to the bed's surface and the vertical (see SI section S2.1 for a detailed derivation). Equation \ref{stress} follows without approximation from the Stokes equation. The first term of Eq. \ref{stress}, $\rho_f g S D$, is simply proportional to the weight of the water column. It corresponds to the stress that the fluid would exert on a perfectly flat surface. It ignores the transfer of momentum across stream and we will call it the ``shallow-water component'', in reference to the celebrated shallow-water approximation. The second term, $\rho_f \nu (\bar{u} D)''$, accounts for the viscous transfer of momentum across the stream (along $y$), and we will call it the ``momentum diffusion component''. Finally, the term $\cos \phi$ accounts for the orientation of the bed surface. Equation \ref{stress} is not closed --- in order to find $\tau$, we still need to solve the Stokes equation for $u$ to get the vertically averaged velocity, $\bar{u}$. Since we hope to bypass the solution of the Stokes equation, Eq. \ref{stress} is not very useful in its present form; we will, however, close it by assuming the river is much wider than it is deep (section \ref{sec:bvp}).

\subsection*{Sediment transport}

If the forces acting to dislodge sediment grains are too weak, the grains remain trapped on the river bed, and there is no sediment transport \cite{shields1936application}. The existence of this threshold force is an instance of Coulomb's law of friction \cite{seizilles2013width}. 

On a flat bed, the fluid acts tangentially to the bed surface, dislodging the grains, while gravity acts normally, anchoring the grains to the bed. In such a case, the sediment flux depends on the so-called Shields parameter, $\theta$, which is proportional to the ratio $F_f/F_g$ of the fluid force acting on a single grain, $F_f \propto \tau d_s^2$, and the grain's weight, $F_g \propto (\rho_s - \rho_f) g d_s^3$ \cite{shields1936application}:
\begin{linenomath*}
\begin{equation} \label{Shields_parameter}
\theta \equiv \frac{\tau}{ (\rho_s - \rho_f) g d_s} \text{  . } 
\end{equation}
\end{linenomath*}
The onset of sediment transport is a complicated phenomenon under active investigation \cite{houssais2015onset,pahtz2020physics,salevan2017determining}. However, a simple representation of sediment transport is to assume that on a flat bed, there exists a threshold Shields parameter, $\theta_t$, below which there is no sediment transport, while for small deviations above this threshold, the sediment flux, $q_s$, increases linearly with the distance to the threshold \cite{lobkovsky2008erosion}, 
\begin{linenomath*}
\begin{equation} \label{flat_bed_transport_law}
q_s = q_0 (\theta - \theta_t) \quad \text{for} \quad \theta > \theta_t \text{  . } 
\end{equation}
\end{linenomath*}
The values of $\theta_t$ and $q_0$ can be directly measured in experiments. The pre-factor $q_0$ is of the order of the ratio of the velocity, $v_s$, of a moving grain to its area, $d_s^2$ --- $q_0 \propto v_s / d_s^2$, where $v_s$ is proportional to the Stokes settling velocity, $v_s \propto (\rho_s - \rho_f) d_s^2 g / \rho_f \nu$ \cite{charru2004erosion,charru2006selection}.

On a rounded bed (as in Figs. \ref{fig:2}a and b), we cannot simply use the Shields parameter as a criterion for grain motion, since gravity has both a normal and a tangential component with respect to the bed surface. Grains in such a configuration begin to move when the ratio, $\mu$, of tangential forces acting to dislodge the grains to normal forces acting to keep them in place becomes greater than a certain value, $\mu_t$, which we can roughly interpret as the friction coefficient \cite{seizilles2013width}. We can estimate this friction coefficient independently from $\theta_t$ in experiments, e.g. by building a heap of sediment and finding the angle at which its grains begin to topple. Abramian et al. \cite{abramian2018self} hypothesized that the transport law for the flat bed can be generalized to a curved bed --- i.e. that the flux, $q_s$, is proportional to the distance of $\mu$ to threshold, $\mu_t$:
\begin{linenomath*}
\begin{equation} \label{qs_mu}
q_s = q_\mu(\mu - \mu_t) \quad \text{for} \quad \mu > \mu_t \text{  . } 
\end{equation}
\end{linenomath*}
To keep this expression consistent with Eq. \ref{flat_bed_transport_law} for the flat bed, we must have $q_\mu \equiv q_0 \theta_t/\mu_t$, since, on a flat bed, $\mu = \mu_t \theta / \theta_t $ \cite{seizilles2013width}. Although Eq. \ref{qs_mu} is difficult to test independently in an experiment, we will show that it is consistent with the experiments of Abramian et al. \cite{abramian2020laboratory}. Parameters $\theta_t$, $\mu_t$, and $q_0$ depend on the grain shape and on the Reynolds number at the grain scale. Abramian et al. \cite{abramian2018self} found them to be $\theta_t = 0.167 \pm 0.003$, $\mu_t = 0.9 \pm 0.2$, and $q_0 = 544 \pm 48 \text{ grains cm}^{-1}\text{s}^{-1}$ in their experiments. 

To find $\mu$, we need to consider the forces acting on a grain of sediment --- the fluid force, $F_f$, acts tangentially, while gravity has both a tangential (downhill) component, $F_g \sin \phi$, and a component normal to the bed, $F_g \cos \phi$. Because the downstream slope of a channel, $S$, is small, the gravitational force is approximately perpendicular to the fluid shear force, $F_f$, and the force ratio, $\mu$, is therefore
\begin{linenomath*}
\begin{equation} \label{mu_def}
\mu = \sqrt{\left(\frac{F_f}{F_g  \cos \phi}\right)^2 +  (\tan \phi) ^2 }\text{  , } 
\end{equation}
\end{linenomath*} 
The ratio of fluid force to gravity, $F_f/F_g$, is proportional to the Shields parameter. In particular, we must have $F_f/F_g = \mu_t \theta / \theta_t$, since, on a flat bed, $\mu = \mu_t$ when $\theta = \theta_t$. With this relation, using Eq. \ref{stress} for stress, and relating $\phi$ to depth as $\tan \phi = D'$, we can express the force ratio $\mu$ from Eq. \ref{mu_def} as 
\begin{linenomath*}
\begin{equation} \label{mu_prelim}
\mu =\sqrt{ \left(\frac{ \mu_t \rho_f S}{ \theta_t (\rho_s - \rho_f) d_s } \right)^2 \left( D + \frac{\nu}{ g S} (\bar{u} D)'' \right)^2 +  D'^2 }\text{  , } 
\end{equation}
\end{linenomath*}
Neglecting the cross-stream momentum diffusion, $(\bar{u} D)''$, yields a purely shallow-water model, which Seizilles et al. \cite{seizilles2013width} used to find the shape of inert rivers.

\subsection*{Sediment diffusion}

Due to random interactions with the river bed, grains traveling downstream also diffuse laterally, towards areas of the bed where sediment transport is less intense \cite{seizilles2014cross}. This cross-stream diffusion of sediment opposes gravity, which pulls the grains down towards the center of the channel. Abramian et al. \cite{abramian2019boltzmann} showed that, in equilibrium, the downhill flux of sediment due to gravity is balanced by this uphill diffusive flux of sediment. Like the Boltzmann equilibrium of a gas in a gravitational field, this balance leads to the exponential distribution of the moving grains as a function of the flow depth:
\begin{linenomath*}
\begin{equation} \label{Boltzmann_simple}
q_s = q_{\mathrm{B}} e^{ D / \lambda} \text{  . } 
\end{equation}
\end{linenomath*}
The last parameter in this equation, $\lambda$, is the characteristic scale for sediment diffusion, and is analogous to the temperature in a gas. Since sediment diffusion is driven by the bed roughness, $\lambda$ scales with the grain size ($\lambda \approx 0.12 d_s \pm 20\% $ \cite{abramian2019boltzmann}).

The prefactor, $q_{\mathrm{B}}$, is the sediment flux at the banks of the river ($D = 0$). Since the flux at the banks is very small compared with the flux at the bottom, $q_{\mathrm{B}}$ does not yield the correct scale for the sediment flux ($q_{s,\mathrm{max}} / q_{\mathrm{B}} \sim 10^{23}$ for the experiment with $Q_s = 60 \text{ grains s}^{-1}$). For this reason, we rewrite Eq. \ref{Boltzmann_simple} in a more convenient form by defining a parameter $\xi$ with units of depth, such that $q_{\mathrm{B}} \equiv q_\mu e^{-\xi / \lambda}$, where $q_\mu$ is the prefactor of the sediment transport law, Eq. \ref{qs_mu}. In this way, $q_\mu$ gives the correct scale for the sediment flux, while $\xi$ is of the order of the maximum depth of an active river ($q_{s,\mathrm{max}} / q_\mu \sim 0.2$ and $\xi/D_\text{max} \sim 1.05$ for the experiment with $Q_s = 60 \text{ grains s}^{-1}$). As we will see below, the maximum river depth, $D_\text{max}$, is generally less than $\xi$, so the maximum flux in a river is typically less than $q_\mu$. With this, Eq. \ref{Boltzmann_simple} becomes
\begin{linenomath*}
\begin{equation} \label{Boltzmann}
q_s = q_\mu e^{( D - \xi ) / \lambda} \text{  . } 
\end{equation}
\end{linenomath*}
The parameter $\xi$ controls the intensity of sediment flux and ensures that the sediment discharge is the integral of the flux, $Q_s =  q_\mu \int \mathrm{exp}(( D - \xi ) / \lambda) \text{d}y$. As an integration constant, its value depends on the discharges transported by the river, but it is not immediately obvious how. A vanishing sediment discharge in rivers corresponds to $\xi \rightarrow \infty$, while finite values of the sediment discharge correspond to smaller values of $\xi$. Sediment transport in a river is significant when the difference, $D_\text{max} - \xi$, between the river depth and $\xi$, is of the order of $\lambda$. This is why $\xi$ of active rivers is of the order of the maximum depth, while it is much greater than the depth of inert ones. In the experiments of Abramian et al. \cite{abramian2020laboratory}, $\xi$ is not set a priori, but only becomes measurable after the river has formed, and, in that sense, plays a similar role as the slope, $S$. 

Equation \ref{Boltzmann} relates the sediment flux, $q_s$, to the river shape, $D(y)$, and has been confirmed repeatedly in experiments \cite{seizilles2014cross,abramian2019boltzmann}. We note that, unlike the gas which simply adjusts to the external field, the river selects its own potential (i.e. its own shape), $D(y)$.

\section{Boundary value problem}\label{sec:bvp}

The relations for the flow, sediment flux, and sediment diffusion we introduced above combine to determine the equilibrium shape of a river. In particular, the Stokes law, Eq. \ref{Stokes_law}, relates the river depth profile, $D(y)$, to the vertically averaged fluid velocity profile, $\bar{u}(y)$. Then, the sediment flux equations, Eqs. \ref{qs_mu} and \ref{mu_prelim}, relate this fluid velocity to the sediment flux profile. Finally, the Boltzmann distribution, Eq. \ref{Boltzmann}, relates the sediment flux back to the depth profile, thereby closing the system of equations for $D(y)$. However, solving these equations simultaneously to get a self-consistent depth and sediment flux profiles is a difficult task --- one needs to solve a two-dimensional, partial differential equation with a moving boundary. Instead, in order to make sense of these equations, we propose to approximate the average velocity, $\bar{u}$.

Seizilles et al. \cite{seizilles2013width} showed that the shallow-water approximation accounts for the equilibrium shape of inert laminar rivers. This approximation, which assumes that there is no transfer of momentum between adjacent fluid columns, i.e. that we can neglect the $y$-derivatives of $u$ in the Stokes equation, is exact when the bed is completely flat. It also works well when depth variations occur on length scales that are much longer than the depth itself. In the case of our river, this would apply when the aspect ratio of the river is large. Neglecting the $y$-derivatives in Eq. \ref{Stokes_law}, we find that the vertically averaged shallow-water velocity, $\bar{u}_{\text{sw}}$, is proportional to the square of the depth, a result known as the lubrication approximation \cite{goodwin1991viscous}
\begin{linenomath*}
\begin{align} \label{u_sw}
\bar{u}_{\text{sw}} =\frac{g S D^2}{3 \nu} \text{ . }
\end{align}
\end{linenomath*}
On a flat bed, where Eq. \ref{u_sw} is exact, the fluid stress, $\tau$, would only contain the shallow-water contribution proportional to depth, $\tau_\text{sw} = \rho_f g S D$ (Eq. \ref{stress}). Approximating stress in this way would allow us to close the system of equations for the river shape, in a way similar to Seizilles et al. \cite{seizilles2013width}. However, it turns out that keeping only the shallow-water contribution to the stress yields unrealistic profiles for active rivers (i.e. when $Q_s > 0$) (SI section S7.1). Parker \cite{parker1978self} first suggested that the cross-stream diffusion of momentum plays an important role for bedload sediment transport in rivers. In line with his suggestion, we keep the momentum diffusion term, $(\bar{u} D )''$, in the expression for the stress, but approximate $\bar{u}$ with the shallow-water velocity, $\bar{u}_{\text{sw}}$, given by Eq. \ref{u_sw}. Then, combining Eqs. \ref{qs_mu}, \ref{mu_prelim}, and \ref{Boltzmann}, we get an ordinary differential equation expressed solely in terms of the depth and its derivatives:
\begin{linenomath*}
\begin{align} \label{BVP}
\sqrt{ \frac{S^2}{L_s^2} \left( D + \tfrac{1}{3} (D^3 )'' \right)^2 + D'^2 } - \mu_t= e^{( D - \xi )/\lambda} \text{  , } 
\end{align}
\end{linenomath*}
where we have introduced a length scale of the order of the grain size, $L_s$, that is a combination of parameters directly measurable in our experimental setup: 
\begin{linenomath*}
\begin{align} \label{Ls}
L_s \equiv \frac{  \theta_t (\rho_s - \rho_f) d_s }{   \mu_t \rho_f } \text{  . }
\end{align}
\end{linenomath*}
We discuss this approximation in detail in the SI sections S2.2 and S2.3, where we show that it is the first term in a series expansion for large aspect ratio, $W/D_\text{max}$ --- it corrects the shallow water stress with a term of order $D_\text{max}^2/W^2$. There, we also show that the contribution of momentum diffusion, $(D^3 )''/3$, in Eq. \ref{BVP} is of the same order as the contribution of gravity, $D'$, so that it should not be neglected in a self-consistent model of a river (SI section S2.2). Recognizing that momentum diffusion is essential to form active rivers, and finding a suitable approximation for it, is a major theoretical contribution of our paper. In principle, Eq. \ref{BVP} could fail to be a meaningful approximation of the stress in a channel with an aspect ratio of order one, but, in our case, it meaningfully corrects the stress for rivers under all experimental conditions we tested (even in the case of inert rivers with $W/D_\text{max} \sim 4$).

Equation \ref{BVP} is an ordinary differential equation. To solve this second-order problem, we need to specify two boundary conditions. For a solution of Eq. \ref{BVP} to be a river, the depth needs to vanish on the banks and the center needs to be flat. Therefore, Eq. \ref{BVP} is a boundary value problem (BVP) with boundary conditions $D(y = - W/2) = 0$ and $D'(y = 0) = 0$. There are several parameters that enter our equation, some of which are directly measurable in our experimental setup ($\mu_t$, $\lambda$, and $L_s$), while others depend implicitly on the discharges of fluid and sediment and become apparent only after the river has formed ($S$ and $\xi$). Although the river width, $W$, is unknown a priori, it is not an independent parameter --- it can be inferred through solving Eq. \ref{BVP} for a given choice of other parameters (SI section S3.1). We emphasize that Eq. \ref{BVP} describes the equilibrium river profile, and, therefore, does not convey anything about transient, time-dependent processes that occur as the river approaches the equilibrium. 

\section{Dependence on water and sediment discharge}\label{sec:dependence_on_discharges}

If we choose the parameters $\mu_t$, $\lambda$, $L_s$, $S$ and $\xi$, we can numerically solve Eq. \ref{BVP} to get a unique river profile, $D(y)$ (SI section S3.1). However, since $S$ and $\xi$ are not directly measurable in our experiment, we cannot immediately determine the shape of the river by simply prescribing the discharge of fluid and sediment in the same way as we would in an experiment. The dependence of $S$ and $\xi$ on the discharges is complicated, and, on the theoretical grounds, we can only say that the inert river, $Q_s = 0$, corresponds to $\xi \rightarrow \infty$, while active rivers correspond to smaller values of $\xi$. Nevertheless, we can find this dependence numerically as follows. For each solution, $D(y)$, of our equation that corresponds to a particular choice of $S$ and $\xi$, we can find the discharges of fluid and sediment as
\begin{linenomath*}
\begin{align} \label{dimensional_from_nondimensional_discharges}
Q_w = \int_{-W/2}^{W/2} \frac{ g S D^3 }{ 3 \nu } \text{d}y \quad \text{,} \quad  Q_s =  \int_{-W/2}^{W/2} q_\mu e^{ ( D - \xi )/\lambda} \text{d}y  \text{  ,}
\end{align}
\end{linenomath*}
where we relate the fluid discharge, $Q_w = \int D \bar{u} \text{d}y$, to the depth profile by approximating $\bar{u}$ with the shallow-water velocity, $\bar{u}_\text{sw}$ (Eq. \ref{u_sw}), and use the Boltzmann distribution (Eq. \ref{Boltzmann}) to relate the sediment flux to depth. Keeping $\mu_t$, $\lambda$, and $L_s$ fixed to their experimental values, $Q_w$ and $Q_s$ are only functions of the parameters $S$ and $\xi$. Inverting these relations numerically yields the model parameters as functions of the discharges of water and sediment, $S(Q_w,Q_s)$ and $\xi(Q_w,Q_s)$. This allows us to directly compare our theory to the experiments (SI section S3.2). We find that the theoretical cross-sections and sediment flux profiles resemble their experimental counterparts, without any fitting parameter (Fig. \ref{fig:2}). 

Encouraged by this result, we now describe how our theoretical rivers depend on $Q_w$ and $Q_s$ (see also SI section S3.3). As we increase the water discharge, $Q_w$, the width and depth of the river increase approximately as $Q_w^{1/3}$, while its slope decreases roughly as $Q_w^{-1/3}$ (SI Fig. S5f), in accordance with the result of Seizilles et al. \cite{seizilles2013width} for inert rivers. This $1/3$ exponent is a signature of the laminar flow in our rivers, in contrast with natural turbulent ones which scale with the $1/2$ exponent of the empirical Lacey's law \cite{savenije2003width,metivier2017laboratory}. Though the size of a river in our model may vary by orders of magnitude under varying $Q_w$, its shape, described for example by the aspect ratio, does not change much unless the river transports a significant amount of sediment. On the other hand, increasing $Q_s$ while keeping $Q_w$ fixed makes the river wider and shallower, while affecting its overall scale only slightly. In short, the water discharge sets the size of the river, while the sediment discharge sets its shape. 

In Fig. \ref{fig:3}, we show that our predictions fall within the uncertainty range of observations of Abramian et al. \cite{abramian2020laboratory}. The aspect ratio and the transport width, $W_T$, increase with sediment discharge in both the model and the experiments (Figs. \ref{fig:3}a and d). The sediment flux increases and saturates for large sediment discharge (Fig. \ref{fig:3}c). This explains why the river becomes wider as we increase $Q_s$ --- if the sediment flux, $q_s$, saturates, the river needs to widen to accommodate a larger sediment discharge. At the same time, this widening forces the river to become shallower in order to maintain a constant fluid discharge, $Q_w$, so that its overall size does not change much while its aspect ratio grows. The simple, nearly linear relationship between the aspect ratio and the sediment discharge shown in Fig. \ref{fig:3}a means that this basic geometric property of the river shape can be used to infer the sediment load, at least in the case of straight, laminar, single thread rivers. 

Since the downstream slope, $S$, is very small, it cannot be measured directly in the experiments. Nevertheless, our theory makes a prediction for it, which we show in Fig. \ref{fig:3}b: the predicted slope is of the order of 0.01, and increases almost linearly with $Q_s$. 

\section{Inert, active, and limiting river}\label{sec:phase_space}

\begin{figure}[t]
\centering
\includegraphics[width=1.\linewidth]{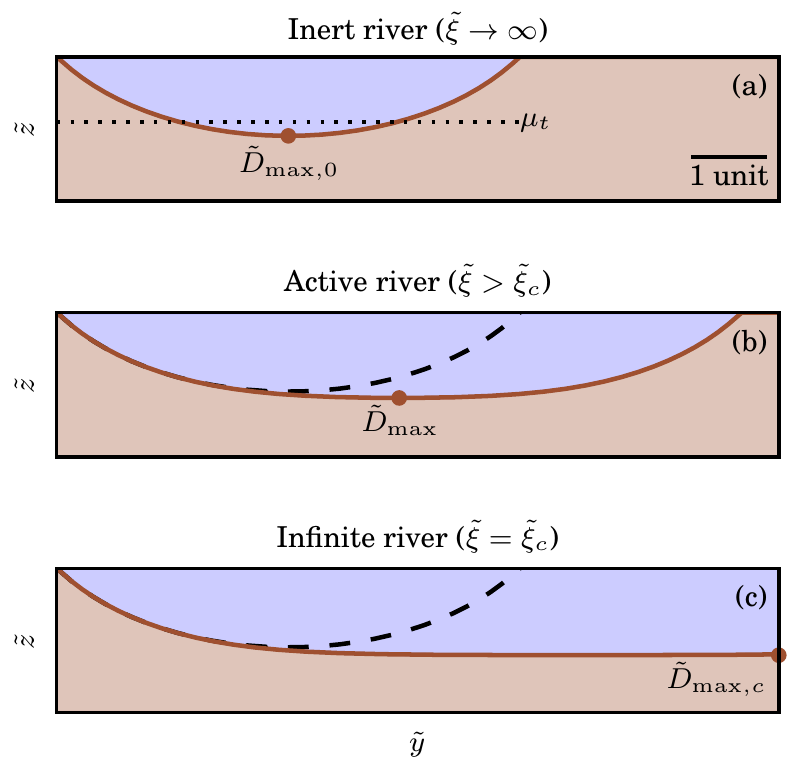}
\caption{River solutions in the non-dimensional $\tilde{y}$-$\tilde{z}$ space for $\mu_t = 0.9$, $\tilde{\lambda} = 0.1$, and varying $\tilde{\xi}$. Black dashed lines in panels (b) and (c) represent the inert river from the upper panel. Aspect ratio is preserved. (a) Inert river ($\tilde{\xi} \rightarrow \infty$). Brown dot marks the non-dimensional inert river depth, $\tilde{D}_{\mathrm{max},0}$. This depth is greater than the friction coefficient, $\tilde{D}_{\mathrm{max},0} > \mu_t$, marked by the horizontal dotted line. (b) Active river ($\tilde{\xi} = 1.33 > \tilde{\xi}_c$). Brown dot marks the depth, $\tilde{D}_{\mathrm{max}}$. (c) Infinite limiting river ($\tilde{\xi} = 1.3237 \approx \tilde{\xi}_c$). Brown dot marks the limiting depth, $\tilde{D}_{\mathrm{max},c}$. }
\label{fig:5}
\end{figure}

We can simplify our model by making Eq. \ref{BVP} non-dimensional, thereby reducing the number of parameters that represent the river. To that end, we rescale all lengths by $L_s / S$
\begin{linenomath*}
\begin{equation} \label{nondim_params}
\tilde{y} \equiv \frac{ y S}{  L_s } \quad \text{ , } \quad  \tilde{D} \equiv \frac{  D S }{ L_s } \quad \text{ , } \quad \tilde{\lambda} \equiv \frac{ \lambda S }{ L_s } \quad \text{ , } \quad  \tilde{\xi} \equiv \frac{  \xi S }{ L_s } \text{ . }
\end{equation}
\end{linenomath*}
In terms of these non-dimensional parameters, Eq. \ref{BVP} becomes:
\begin{linenomath*}
\begin{align} \label{BVP_nondim}
\sqrt{ \left( \tilde{D} + \tfrac{1}{3} (\tilde{D}^3 )'' \right)^2 +  \tilde{D}'^2 } - \mu_t= e^{( \tilde{D} - \tilde{\xi} ) / \tilde{\lambda}} \text{  , }
\end{align}
\end{linenomath*}
where, now, the primes stand for derivatives with respect to $\tilde{y}$. The non-dimensional depth, $\tilde{D}$, is of order one, regardless of the size of the original river. Therefore, Eq. \ref{BVP_nondim} describes the river shape, while the ratio $L_s / S$ sets its size. The river shape depends on only three non-dimensional parameters  --- $\mu_t$, $\tilde{\lambda}$, and $\tilde{\xi}$. 

When $\tilde{\xi} \rightarrow \infty$ (Fig. \ref{fig:5}a, SI section S4.3), the river becomes inert as the exponential on the right-hand side of Eq. \ref{BVP_nondim} vanishes. In this case, the dependence on $\tilde{\xi}$ and $\tilde{\lambda}$ vanishes, so the river shape depends only on the friction coefficient, $\mu_t$. Since the friction coefficient is a fixed property of the sediment grains, the fluid discharge, $Q_w$, cannot change the shape of such a river. Instead, the fluid discharge can only affect its size by changing the scale factor, $L_s / S$. Physically, this is because the laminar flow is scale-free, which makes the inert river shape independent of its size. An inert river, thus, reaches a maximum depth $\tilde{D}_{\text{max},0}(\mu_t)$. To calculate its value, we need to numerically solve Eq. \ref{BVP_nondim}, but unlike the complete theory of section \ref{sec:bvp}, this is a straightforward problem since it depends on a single, directly measurable parameter. For $\mu_t = 0.9$ which corresponds to the experiments, we numerically find $\tilde{D}_{\text{max},0}(\mu_t) \approx 1.1$. From Eq. \ref{BVP_nondim}, we can express this depth as $\tilde{D}_{\text{max},0} = \mu_t - \tfrac{1}{3} (\tilde{D}^3 )''|_\text{center}$, since the gravity contribution, $\tilde{D}'$, vanishes at the river center. Thus, the momentum diffusion contribution to the stress (the term $\frac{1}{3}(\tilde{D}^3)''$ in Eq. \ref{BVP_nondim}) ensures that the dimensionless inert river depth is greater than the friction coefficient ($\tilde{D}_{\text{max},0} > \mu_t$) --- had we ignored the momentum diffusion, the inert river depth would have been exactly $\mu_t$. Physically, the diffusion of momentum relieves some of the stress from the river center, so, to remain at the threshold of sediment motion, the river has to be deeper than it would be without momentum diffusion. In the next section, we will show that this fact is crucial for the transport of sediment. 

As $\tilde{\xi}$ decreases to finite values, the river becomes active (Fig. \ref{fig:5}b). The banks of such a river largely retain the shape of the inert one, but its bottom part, which carries most of the sediment, widens, and the discharge of sediment increases. For a particular value of $\tilde{\xi}$, say $\tilde{\xi}_c$, which depends on $\mu_t$ and $\tilde{\lambda}$, the river becomes infinitely wide and transports an infinite amount of sediment (Fig. \ref{fig:5}c, SI section S4.2). Such a river has a finite, well-defined depth, $\tilde{D}_{\text{max},c}(\mu_t, \tilde{\lambda})$. This means that, for given values of $\tilde{\lambda}$ and $\mu_t$, there exists a river-solution with a highest possible sediment flux, $q_{s,c} = q_\mu ( \tilde{D}_{\text{max},c} - \mu_t )$. The existence of this limiting flux explains the saturation of $q_s$ for large values of total sediment discharge, $Q_s$, that we see in Fig. \ref{fig:3}c. It also means that, in our model, the distance to threshold in a river, $\mu-\mu_t$, is always less than $\tilde{D}_{\text{max},c} - \mu_t$. Numerically, we find $\tilde{D}_{\text{max},c} - \mu_t \approx 0.22 $ for experimental parameters ($\mu_t = 0.9$ and $\tilde{\lambda} = 0.02$). In the next section, we will estimate the limiting flux, $q_{s,c}$, by assuming sediment diffusion is weak ($\lambda \rightarrow 0$), in which case $q_{s,c}$ only depends on the friction coefficient, $\mu_t$.

\begin{figure*}
\centering
\includegraphics[width=1.\textwidth]{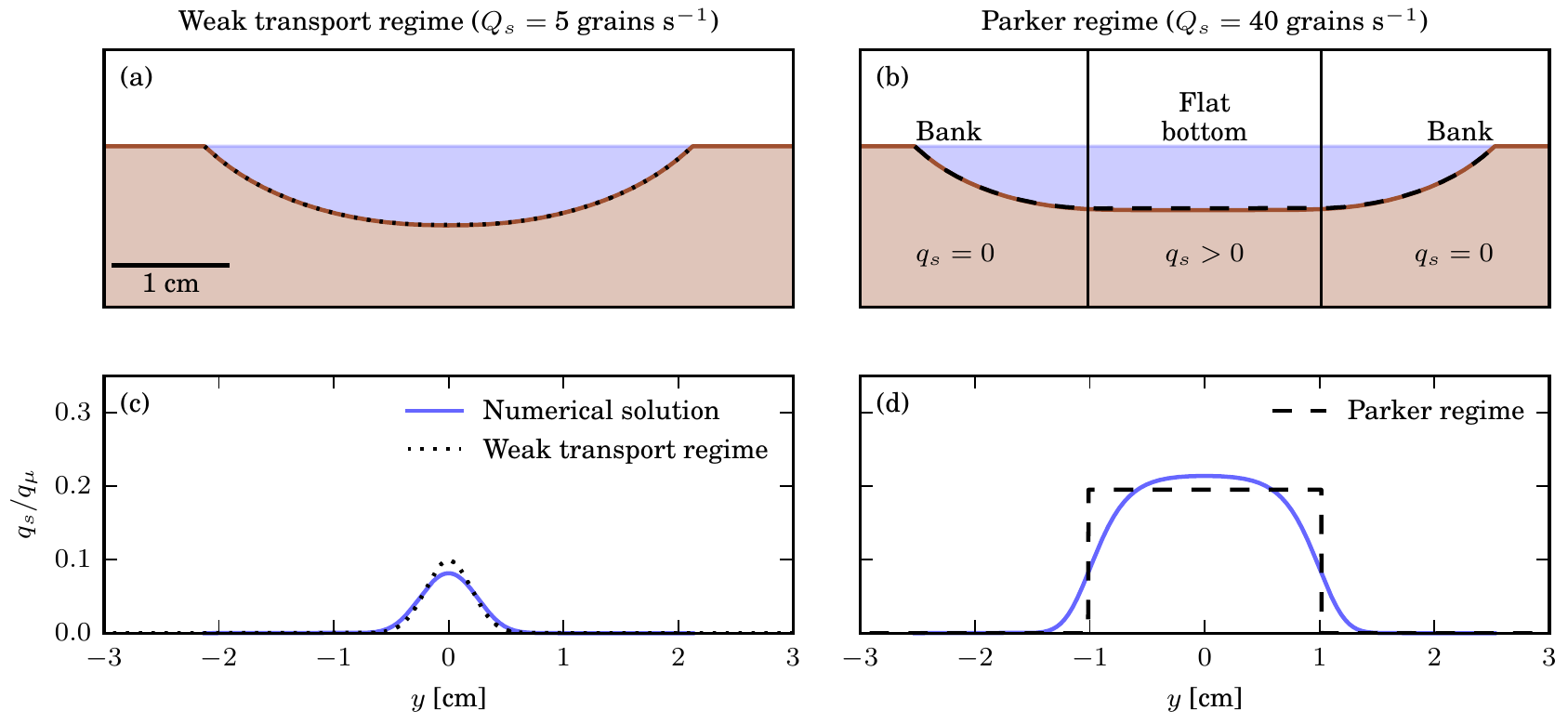}
\caption{Asymptotic regimes. Black dashed lines correspond to the Parker regime while the black dotted lines correspond to the weak transport regime. Top row panels ((a) and (b)) show river depth profiles in our model. The brown lines are numerical solutions of Eq. \ref{BVP} for two values of sediment discharge, $Q_s$. The numerical and asymptotic depth profiles are not distinguishable by eye and the error is of the order $\lambda$ (or about 2\% of the maximum depth) --- the maximum deviation of the asymptotic approximation from the numerical solution is about $0.05 \text{ mm} \approx 0.5 \lambda$ in panel (a) and about $0.15 \text{ mm} \approx 1.5 \lambda$ in panel (b). The vertical black lines in panel (b) split the river in three parts used to construct the Parker river. Lower row panels ((c) and (d)) show the dimensionless sediment flux profiles, $q_s / q_\mu$, that correspond to upper panels. Blue lines are numerical solutions. }
\label{fig:6}
\end{figure*}

\section{The Parker regime}\label{sec:Parker}

Moving grains accumulate at the bottom of the river due to gravity, while they climb back onto the banks by diffusion \cite{abramian2019boltzmann}. The Boltzmann distribution, Eq. \ref{Boltzmann}, implies that diffusion can pull the grains up by a height that is of the order of the length scale $\lambda$. Therefore, the region of the bed over which transport occurs has a depth that is within several $\lambda$ of the maximum, $D_\text{max}$. Since $\lambda$ is small (less than the grain size), rivers that transport a significant amount of sediment need a wide, and essentially flat bottom. Moreover, a small $\lambda$ means the sediment transport decreases rapidly towards the banks, so the banks are nearly inert and, thus, close to the threshold of sediment motion. If, following this reasoning, we neglect sediment diffusion altogether by taking the limit $\lambda \rightarrow 0$, the river sharply separates into a flat, active bottom and curved, inert banks (Fig. \ref{fig:6}b). We will call this simplified configuration the ``Parker regime,'' after Parker \cite{parker1978self} who constructed a similar model for natural gravel-bed rivers. The limit $\lambda \rightarrow 0$ is equivalent to assuming that the fluid discharge is large (to ensure that $\lambda$ is small compared with the width of the inert banks), and that the sediment discharge is large (to ensure that $\lambda$ is small compared with the width of the active, flat bottom). 

We begin the investigation of this regime by first finding the depth, $D_{\text{max}}^{(P)}$, of a Parker river (denoted by the superscript $(P)$). The banks in this approximation are inert and, thus, satisfy our model, Eq. \ref{BVP_nondim}, with the right-hand side set to zero. Therefore, their non-dimensional depth matches that of an inert river, $\tilde{D}_{\text{max},0}(\mu_t)$, and we can set, $\tilde{D}_{\text{max}}^{(P)} = \tilde{D}_{\text{max},0}(\mu_t)$. In dimensional units, this becomes
\begin{linenomath*}
\begin{align} \label{Parker_depth}
D_{\text{max}}^{(P)} = \frac{L_s \tilde{D}_{\text{max},0}(\mu_t)}{S^{(P)}}  \text{  . }
\end{align}
\end{linenomath*}
We note that this dimensional depth of a Parker river differs from that of an inert river, since the slope of a Parker river, $S^{(P)}$, is different from the slope of an inert river, $S_0$ --- these slopes depend on the shape of the entire channel, not only on the banks. The non-dimensional inert river depth, $\tilde{D}_{\text{max},0}$, depends only on the friction coefficient, $\mu_t$ (section \ref{sec:phase_space}). For this reason, the river depth given by Eq. \ref{Parker_depth} is inversely proportional to its slope, $D_{\text{max}}^{(P)} \propto 1/S^{(P)}$, regardless of the fluid and sediment discharges. This is consistent with the original model of Parker, as well as with observations in natural rivers \cite{parker1978self,andrews1984bed}. 

Once we know the depth of a Parker river, we can find its sediment flux. Since the bottom is flat, the cross-sectional profile of the sediment flux is a rectangle of height $q_{s}^{(P)}$ and width $W_T^{(P)}$ (black dashed line in Fig. \ref{fig:6}d). The flat bottom feels only the shallow-water component of the stress so the force ratio on the bottom is $\mu = \tilde{D}_\text{max}^{(P)}$ (Eq. \ref{mu_prelim}). According to the transport law, Eq. \ref{qs_mu}, this yields a sediment flux 
\begin{linenomath*}
\begin{align} \label{Parker_qs}
\frac{q_{s}^{(P)}}{ q_\mu} = \tilde{D}_{\text{max},0}(\mu_t) - \mu_t \text{  , }
\end{align}
\end{linenomath*}
where we used $\tilde{D}_\text{max}^{(P)} = \tilde{D}_{\text{max},0}(\mu_t)$. Therefore, the sediment flux, and, correspondingly, the distance to threshold, $\mu - \mu_t$, depend only on the friction coefficient, $\mu_t$, and have the same value regardless of the discharges of fluid and sediment --- this is the gist of the Parker regime. Numerically solving Eq. \ref{BVP_nondim} for an inert river (with $\mu_t = 0.9$), we find $\tilde{D}_{\text{max},0} - \mu_t \approx 0.2$. The sediment flux of a Parker river, $q_{s}^{(P)}$, is an approximation of the limiting flux, $q_{s,c}$, we discussed in section \ref{sec:phase_space} --- in fact, $q_{s}^{(P)}$ is the limit of $q_{s,c}$ as $\lambda \rightarrow 0$. For typical experimental parameters ($\mu_t = 0.9$ and $\tilde{\lambda} = 0.02$), $q_{s,c}$ is less than 10\% higher than $q_{s}^{(P)}$, so the Parker regime approximates the numerical solution well. 

The sediment flux $q_{s}^{(P)} \approx 0.2 q_\mu$ we find for a laminar river corresponds to a fluid-induced stress on the river bottom that is about $22\%$ higher than critical. Interestingly, this value of the stress is comparable to observations in natural rivers, and to the original Parker's theory for turbulent rivers \cite{dunne2020sets,parker1978self}. There is no reason to expect that this proportion should be exactly the same for laminar and turbulent flows. However, that it is independent from the water and sediment discharges, and of order one, is likely not a coincidence. The scale-independence of the flow ensures that the bank shape (i.e. $\tilde{D}_{\text{max},0}(\mu_t)$) is independent of the discharges (section \ref{sec:phase_space}). Thus, the discharge-independent sediment flux likely results from the scale-independence of the flow, under both laminar and turbulent conditions.

The difference $\tilde{D}_{\text{max},0} - \mu_t$ vanishes in the classical shallow-water approximation (section \ref{sec:phase_space}). As a consequence, if we ignored momentum diffusion, the river could not carry any sediment --- sediment transport in the Parker regime is only possible because momentum diffuses across the stream. Indeed, this tends to homogenize the distribution of shear stress over the bed, especially on the banks, where the bed is curved. As a consequence, the deeper parts of the banks hand over some of the momentum to the shallower parts of the banks. This means that, in order to stay at the threshold of sediment motion, the banks need to be deeper than they would be in the absence of momentum diffusion. This increased depth then causes excess stress on the flat river bottom, which only feels the shallow-water component of the stress, thereby driving sediment transport (SI section S7). This is why we need to keep track of momentum diffusion, even in a minimal model of a river. 

The weakness of sediment diffusion, characterized by a small diffusion length, $\lambda$, ensures that rivers remain close to the threshold. For a small but finite $\lambda$, we numerically find that the maximal distance to threshold is approximately $\tilde{D}_{\text{max},c} - \mu_t \approx \tilde{D}_{\text{max},0}(\mu_t) - \mu_t + S \lambda / L_s$ (SI section S4.4). The term $S \lambda / L_s$ is negligible for large rivers with a small slope, such as the ones in the experiments we are considering ($S \lambda / L_s \approx 0.022$ for highest experimental $Q_s$). The fact that rivers tend towards the Parker regime as the fluid and sediment discharges increase is, thus, the reason they do not exceed the threshold significantly more than $\tilde{D}_{\text{max},0}(\mu_t) - \mu_t$. In short, it is the combination of significant momentum diffusion with weak bedload diffusion that maintains the laboratory rivers of Abramian et al. \cite{abramian2020laboratory} near the threshold.

Once we have identified the sediment flux, $q_s^{(P)}$, all other properties follow straightforwardly. In particular, we can get the width of the active bottom, $W_T^{(P)}$, as 
\begin{linenomath*}
\begin{align} \label{Parker_WT}
W_T^{(P)} = Q_s / q_s^{(P)} \text{  . }
\end{align}
\end{linenomath*}
The total width of a Parker river, $W^{(P)} = W_T^{(P)} + W_0^{(P)}$, is then the sum of $W_T^{(P)}$ and the bank width, $W^{(P)}_0 = \tilde{W}_0 L_s / S^{(P)}$. Here, $\tilde{W}_0$ is the non-dimensional width of an inert river that is only a function of $\mu_t$ (numerically, we find $\tilde{W}_0 \approx 6.4$ for $\mu_t = 0.9$). 

From here, we can find the aspect ratio of a Parker river as 
\begin{linenomath*}
\begin{align} \label{Parker_aspect_ratio}
\frac{W^{(P)}}{D_\mathrm{max}^{(P)}} = \frac{\tilde{W}_0}{\tilde{D}_{\mathrm{max},0}} + \frac{Q_s}{q_s^{(P)} D_\text{max}^{(P)}} \text{  . }
\end{align}
\end{linenomath*}
This equation shows how the geometry of a river can be used to infer its sediment load. Namely, from Eq. \ref{Parker_aspect_ratio}, the sediment discharge is 
\begin{linenomath*}
\begin{align} \label{Parker_sediment_discharge}
Q_s = q_s^{(P)} D_\text{max}^{(P)} \left( \frac{W^{(P)}}{D_\mathrm{max}^{(P)}} - \frac{\tilde{W}_0}{\tilde{D}_{\mathrm{max},0}} \right) \text{  . }
\end{align}
\end{linenomath*}
The quantities $q_s^{(P)}$ and $\tilde{W}_0/\tilde{D}_{\text{max},0}$ are universal in that they only depend on the properties of the sediment and the general properties of the flow (such as its laminarity). As such, they are independent of the discharges of fluid and sediment. All other quantities on the right hand side of Eq. \ref{Parker_sediment_discharge} are geometric ($D_\text{max}^{(P)}$ and $W^{(P)}$). Therefore, one can estimate the sediment load of a river in the Parker regime by simply measuring its width and depth. Equation \ref{Parker_sediment_discharge} follows from general considerations that allow the Parker regime to exist --- such as, for example, that the river splits into an active bottom and inert banks whose shape is independent of the discharges. It is likely that these conditions also apply to turbulent rivers. So, we speculate that Eq. \ref{Parker_sediment_discharge} holds for natural rivers in the Parker regime, although with different values of $q_s^{(P)}$ and $\tilde{W}_0/\tilde{D}_{\mathrm{max},0}$. On an ensemble of rivers with varying fluid and sediment discharge, $q_s^{(P)}$ would represent the maximum observed sediment flux, while $\tilde{W}_0/\tilde{D}_{\mathrm{max},0}$ would be the minimum observed aspect ratio. So, to estimate the sediment load of a natural river, one could begin by estimating the minimum aspect ratio and maximum sediment flux on a large dataset of rivers, and then measuring the width and depth of a particular river. We demonstrate the validity of this method on our experimental dataset in the SI section S6.

The transport width of a Parker river becomes comparable to the river size when the sediment discharge is $Q_s \approx q_s^{(P)} L_s / S^{(P)}$. This defines a characteristic discharge in the Parker regime, $Q_s^*$:
\begin{linenomath*}
\begin{align} \label{Parker_Qs_characteristic}
Q_s^* = q_\mu \left( \frac{\nu Q_w}{g L_s} \right)^{1/3} \text{  , }
\end{align}
\end{linenomath*}
where we approximated the slope with that of an inert river (Eq. S38 of the SI) and neglected dimensionless factors of order one. When the sediment discharge is much greater than $Q_s^*$, the aspect ratio of a river grows with the sediment discharge, while for $Q_s$ much smaller than $Q_s^*$, it becomes that of the inert river. In the experiments, we find $Q_s^* \approx 74 \text{ grains s}^{-1}$. Interestingly, this value is close to the discharge $Q_s \approx 90 \text{ grains s}^{-1}$ at which the experimental rivers destabilize into braids, suggesting that $Q_s^*$ may define an upper bound for the sediment load a single channel can carry. Determining this would, however, require an analysis that is beyond the scope of this paper.

Finally, to find the slope of the Parker river, $S^{(P)}$, we first compute its water discharge, $Q_w$ (Eq.  \ref{dimensional_from_nondimensional_discharges}). The water discharge of a Parker river is a sum of the the bank and the flat bottom contributions. In particular, we find (SI section S5)
\begin{linenomath*}
\begin{align} \label{Parker_L}
Q_w = \frac{ g L_s^4 }{ \nu S^{(P) 3} } \left( \tilde{Q}_{w,0} + \frac{Q_s  S^{(P)} \tilde{D}_{\text{max},0}^3 }{3 q_s^{(P)} L_s}  \right)
\end{align}
\end{linenomath*}
where $\tilde{Q}_{w,0} \equiv \frac{1}{3} \int_{-\tilde{W}_0/2}^{\tilde{W}_0/2} \tilde{D}_0^3 \text{d}\tilde{y}$, is the dimensionless discharge of an inert river with a depth profile $\tilde{D}_0(\tilde{y})$, and is, again, only a function of $\mu_t$ ($\tilde{Q}_{w,0} \approx 1.22$ for $\mu_t = 0.9$). The above equation can be inverted to get $S^{(P)}$ as a function of $Q_w$, $Q_s$, and other measurable parameters, but, since the inverted expression is cumbersome, we do not show it here. To leading order, $S^{(P)} \propto Q_w^{-1/3}$, so the Parker river inherits the basic scaling of laminar rivers \cite{seizilles2013width}.

In Figs. \ref{fig:6}b and d, we show that the cross-section and sediment flux profiles of rivers in the Parker regime capture well the numerical solutions of our model, Eq. \ref{BVP}, when the sediment discharge is large. In Fig. \ref{fig:3}, we compare the Parker river properties to numerical solutions of the full theory and experiments (black dashed lines in Fig. \ref{fig:3}). The slope and shape of numerical solutions are well approximated by the Parker regime for the entire range of sediment discharge (Figs. \ref{fig:3}a and b). Conversely, the sediment flux profile ($q_s$ and $W_T$) for the Parker river is a good approximation of the full theory only when the sediment discharge is large enough. This is not surprising, since, according to the Boltzmann distribution, Eq. \ref{Boltzmann}, the flux is a sensitive function of the depth so, to get a reasonable estimate of the flux, we need to estimate the depth accurately with a precision that is of the order of the diffusion length, $\lambda$. 

When the sediment discharge is small, sediment diffusion becomes important, and the Parker regime cannot account for the sediment transport (Fig. \ref{fig:6}c). In the next section, we consider this weak transport regime. 

\section{Weak transport regime}\label{sec:weak_transport}

When the sediment discharge is small, the sediment flux concentrates about the center of the river, and does not significantly alter its shape. The sediment flux profile in this case is analogous to the density of an ideal gas in a fixed potential --- the fixed bed shape sets the potential well in which the traveling grains distribute themselves. 

A random walker that makes steps of length $\lambda$ in a fixed potential well with a characteristic size $L$ would spend the majority of its time moving around in an area with a size of the order of $\sqrt{\lambda L}$. Therefore, we expect the sediment grains in this weak transport regime to concentrate in a region of a size $W_T \sim \sqrt{\lambda L_s/S_0}$, where $L_s/S_0$ is the characteristic size of an inert river. As a consequence, the sediment flux would be about $q_s \sim Q_s \sqrt{S_0/\lambda L_s}$. Thus, unlike the Parker river which changes its width to accommodate its sediment load, the weak-transport river adjusts its sediment flux.

We can formalize this argument by first assuming that the depth profile is approximately that of an inert river, $D_0$. Close enough to the center, we can approximate this depth with a parabola, 
\begin{linenomath*}
\begin{align} \label{quadratic_approx}
D_0 \approx D_{\text{max},0} - \frac{\kappa}{2} y^2 \text{  , }
\end{align}
\end{linenomath*}
where $\kappa \equiv -D_0''|_\mathrm{center}$ is the curvature of the bed at the center. If only this quadratic part of the depth profile is relevant, the Boltzmann distribution of traveling grains, Eq. \ref{Boltzmann}, becomes a Gaussian:
\begin{linenomath*}
\begin{align} \label{Gauss_qs}
q_s = q_{s,\text{max}} e^{-\frac{\kappa}{2 \lambda} y^2} \text{  . }
\end{align}
\end{linenomath*}
Here, $q_{s,\text{max}}$ is a constant that depends on $q_\mu$, $\lambda$, and $\xi$. This approximation is valid when the sediment discharge is small enough to leave the depth profile unaltered, and when the fluid discharge is large enough to keep $\sqrt{\lambda / \kappa}$ small compared with the river size.

To specify the sediment flux profile, we first relate the curvature at the river bottom, $\kappa$, to the depth of an inert river, $D_{\text{max},0}$, using Eq. \ref{BVP}. Then, by integrating the sediment flux profile, Eq. \ref{Gauss_qs}, we can find the maximum flux, $q_{s,\text{max}}$, and the transport width, $W_T = Q_s /\langle q_s \rangle$, as functions of $Q_s$ (we find $\langle q_s \rangle$ through Eq. \ref{mean_qs_def}):
\begin{linenomath*}
\begin{align} 
& q_{s,\text{max}}=  Q_s  \sqrt{\frac{ S_0 (\tilde{D}_{\text{max},0} - \mu_t)}{2 \pi \lambda L_s \tilde{D}_{\text{max},0}^2} } \text{  , } \label{Gauss_qsmax} \\
& W_T =  \sqrt{\frac{ 4 \pi  \lambda L_s \tilde{D}_{\text{max},0}^2 }{S_0(\tilde{D}_{\text{max},0} - \mu_t)}} \text{  , } \label{Gauss_WT}
\end{align}
\end{linenomath*}
where $S_0$ can be estimated from the fluid discharge, Eq. \ref{dimensional_from_nondimensional_discharges}, using the inert river profile (Eq. S38 of the SI). We can see that $q_{s,\text{max}} \propto Q_s \sqrt{S_0/\lambda L_s}$ and $W_T \propto \sqrt{\lambda L_s/S_0}$, as anticipated.

Unlike the Parker regime, the weak transport regime requires sediment diffusion --- it does not exist when $\lambda$ vanishes. Figures \ref{fig:3} and \ref{fig:6} show that the sediment flux profile in the numerical model transitions smoothly from the weak transport regime to the Parker regime. This transition happens when $Q_s$ approximately equals $q_\mu \sqrt{\lambda L_s/S_0}$, at which point the weak transport sediment flux overcomes the limiting flux of the Parker regime. This defines a transitional sediment discharge, $Q_{s,\text{t}}$, given by
\begin{linenomath*}
\begin{align} 
&Q_{s,\text{t}} =  q_\mu \left( \frac{\lambda^3 \nu Q_w}{g L_s} \right)^{1/6} \text{  . } \label{Qs_transition}
\end{align}
\end{linenomath*}
For the experiments we are considering, we find $Q_{s,\text{t}} \approx 8.6 \text{ grains s}^{-1}$. A large river (with $\lambda S / L_s \rightarrow 0$) remains in the weak transport regime when $Q_s \ll Q_{s,\text{t}}$, and enters the Parker regime if $Q_s \gg Q_{s,\text{t}}$. When $\lambda$ vanishes, the transitional discharge tends to zero, and the river is always in the Parker regime. Interestingly, the experiments span both regimes --- one experimental run of Abramian et al. \cite{abramian2020laboratory} has $Q_s \approx 12.6 \text{ grains s}^{-1}$, comparable to $Q_{s,\text{t}}$. This means that $\lambda$ in the experiments is small enough for the two regimes to be valid approximations, but still large enough for the weak transport regime to be visible. 

\section{Conclusions}

In this paper, we relied on a recent experimental success in obtaining single-thread laminar rivers, and we developed a physical theory that correctly represents the shape of a river as a function of its water and sediment discharges. In steady-state, the balance between gravity and the stress induced by the fluid flow, along with the diffusion of sediment across the channel, determines the shape of the river. We greatly simplify the problem of calculating the fluid stress, which in general depends on the entire channel shape, by relating it to the local river depth and its derivatives only. Although this model can be numerically solved relatively easily, the relationship between the river properties and the discharges of water and sediment is not immediately obvious. Fortunately, when the sediment discharge is small or large, the relationships between the properties of the river and its discharges reduce to simple algebraic expressions. When the sediment load of a river is large, we find that the diffusion of momentum across the stream generates an excess of stress on the river bottom, which drives sediment transport. Momentum diffusion, thus, plays a key role in determining the shape of the channel, in accordance with the model originally proposed by Parker \cite{parker1978self}. This is not the case in the weak transport regime, which relies on the diffusion of sediment.

Rivers in our model never deviate much from the threshold of sediment motion. As their sediment discharge increases, the sediment flux approaches a maximum, which forces the river to widen and get shallower. This saturation of the sediment flux results from the weakness of sediment diffusion and the scale-independence of the flow. Most likely, natural rivers also meet these conditions, which is why the original model of Parker \cite{parker1978self} has proven to be a fair representation of natural, gravel-bed rivers. According to our theory, however, there exists another, small discharge regime in which a river's shape is independent of its sediment discharge, while its sediment flux is proportional to it. To our knowledge, this regime remains to be identified in the field.

Although our model is aimed at a relatively narrow subset of rivers (straight, laminar rivers that transport uniform, non-cohesive sediment as bedload and with constant water and sediment discharge), it is tempting to extrapolate it to natural rivers which have been observed to maintain their channel close to the threshold of sediment motion. The common explanation for this is that hillslope processes, which feed rivers with sediment, are slow, so that rivers carry only a small sediment load. We show that this is not necessarily true --- our rivers are always close to the threshold, regardless of the sediment discharge, due to the nature of the fluid-sediment interaction. 

Increasing the sediment discharge significantly beyond the last experimental point of Abramian et al. \cite{abramian2020laboratory} destabilizes the rivers into several smaller channels that form a braided river. Curiously, this happens when the transport width of a river becomes comparable to its size. In the future, this may help us identify a mechanism for braiding, which is still debated \cite{leopold1957river,parker1976cause,seminara2010fluvial}.

Our model provides a link between the shape of the river and its sediment load. It, thus, presents an opportunity for field measurements, whereby one could estimate the sediment discharge of a river by measuring its width and depth. Before this method can bed applied to natural rivers reliably, we should first extend the present theory to the case of turbulent flow, which will be the focus of future work.

\textbf{Acknowledgements:} PP was supported by the JSMF postdoctoral fellowship. OD was partially funded by the ``{\it Emergence en Recherche}'' grant of the {\it Universit\'e de Paris}.

\bibliographystyle{apsrev4-1}
\bibliography{bibRivers}{}

\end{document}



\maketitle

\section{Experimental parameters and uncertainties}

In this section we summarize the parameters used in the experiments of \citet{abramian2020laboratory}, show the results for each of the experimental runs, and describe how we estimate the river properties and their uncertainties. 

\subsection{Details of the experimental runs}

In Table \ref{tab:1}, we show the experimental parameters and their uncertainties (uncertainties are estimated in \citet{abramian2020laboratory}). In Fig. \ref{all_profiles}, we show the depth and sediment flux profiles for each of the experiments of \citet{abramian2020laboratory} and compare them to our model predictions. In Table \ref{tab:2} and Fig. \ref{fig:compare_dimensional_properties}, we show the properties of these rivers. 

We can see that our model predicts rivers that are slightly wider than the experiments (by about 20\%), while the depth, $D_\text{max}$, is not biased in an obvious way. The width of the sediment flux profiles is captured quite accurately, but the experimental profiles seem to be somewhat higher (by about 30\% for the maximum sediment flux). 

In our model, the bed slope at the river bank approximately equals the friction coefficient, $D'|_{D=0} \approx \mu_t$, since the stress, $\tau$, vanishes when $D = 0$ and the sediment flux, $q_\mu \text{exp}[-\xi/\lambda]$, is negligible (Eq. 13 of the main text). However, some of the experiments show a high bank angle, unusual for granular material with irregular grains. This could be due to capillary forces acting near the bank, to rivers being not fully in equilibrium, to surface armoring by removing the loose grains \cite{charru2004erosion}, or to the fact that the fluid stress does not necessarily vanish near the bank. In fact, using the value $\mu_t = 1.2$, combined with a length-scale, $L_s$, smaller by about 20\% (which is within the experimental uncertainty), makes our model agree with the experiment well in all metrics. However, to avoid treating $\mu_t$ as a tuning parameter, we decided to use $\mu_t = 0.9$, which is the largest value estimated in an independent experiment \cite{abramian2018self}. 

\subsection{Estimating the river properties and their uncertainties}

\begin{figure}
\centering
\includegraphics[width=\textwidth]{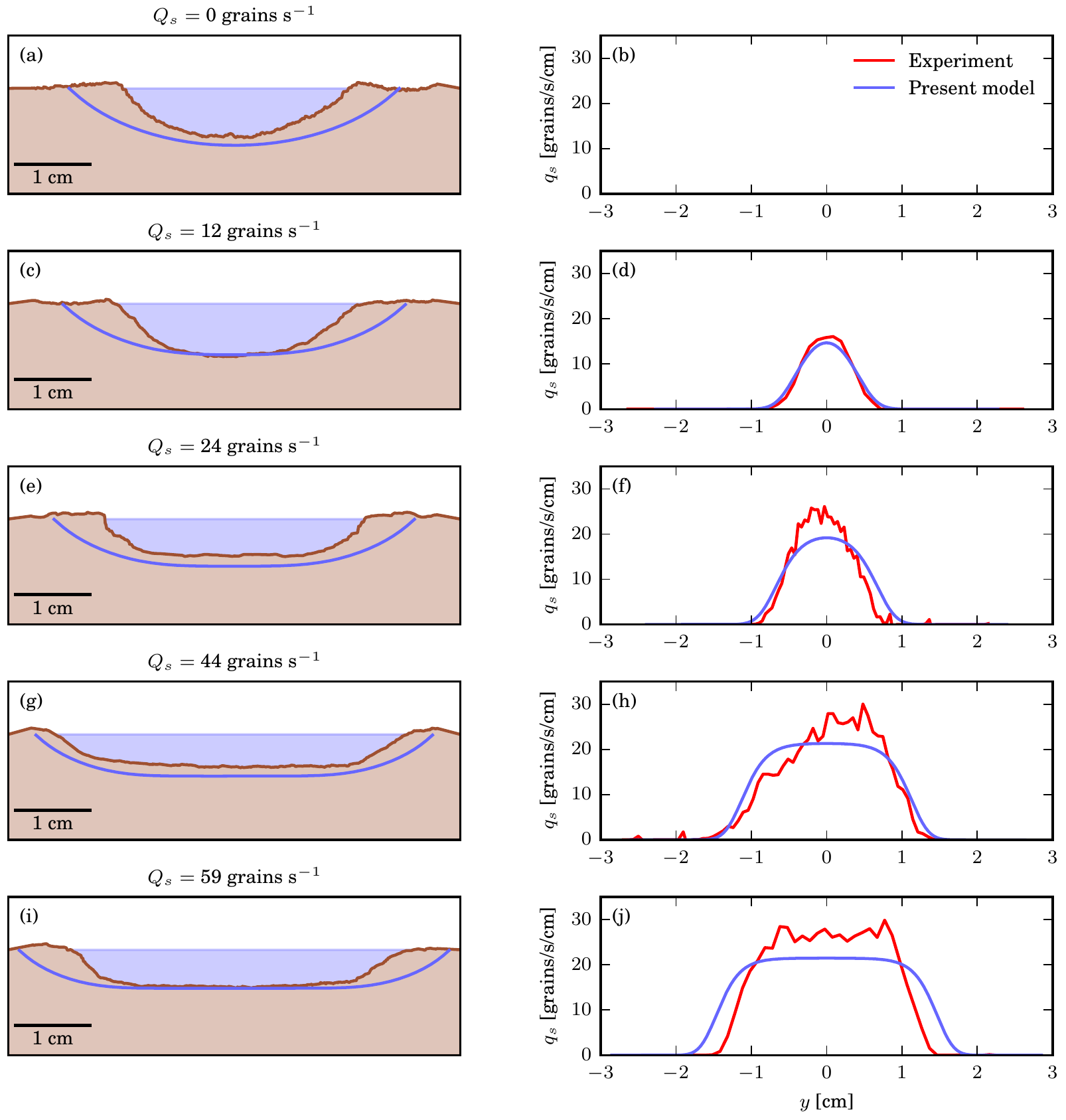}
\caption{Depth and sediment profiles for all experiments of \citet{abramian2020laboratory}. The left column are the measured (brown lines) and modeled (blue dashed lines) depth profiles. The right column are the measured (red lines) and modeled (blue lines) sediment flux profiles.}
\label{all_profiles}
\end{figure}

We estimated all of the river properties based on the cross-sections of Fig. \ref{all_profiles}. In particular, we estimated the total sediment discharge, $Q_s$, as the integral of the sediment flux profile found by grain tracking. We identify the uncertainty in the sediment discharge from the fluctuations of $Q_s$ about the mean, once the equilibrium is reached (Fig. 2 in \citet{abramian2020laboratory}). Next, we estimate the width, $W$, as the distance from one bank to the other, where we identify the banks by the sudden change of slope. In doing so, we introduce an error that is of the order of the grain size. However, we can also estimate the natural variability of the river width along its path from the overhead images of the experiment. We find it to be about $5\%$ of the mean --- an error larger than that introduced by the cross-section measurement. For this reason, we identify the uncertainty for the width with this natural variability. Next, we estimate the depth, $D_\text{max}$, by taking the minimum of a parabola we fit to the the river bottom, in order to minimize the error due to bed roughness. \citet{abramian2020laboratory} estimate the measurement error of depth due to laser inaccuracy to be about $0.5 \text{ mm}$. We do not have access to the natural variability of the depth, although we can say that it is at least of the order of the grain size. Again, this natural variability is greater than the measurement error estimated by \citet{abramian2020laboratory}, so we take the error of $D_\text{max}$ to be $d_s = 0.83 \text{ mm}$ for all experiments. We follow a similar procedure to find the maximum sediment flux, $q_{s,\text{max}}$ --- we fit a parabola around the center of the sediment flux profile, and take its maximum to be $q_{s,\text{max}}$. Again, we do not have access to the natural variability of $q_{s,\text{max}}$ along the river's path, so we take the error of $q_{s,\text{max}}$ to be the difference between the observed maximum sediment flux profile and the height of the fitted parabola. 
 
\begin{figure}
\centering
\includegraphics[width=\textwidth]{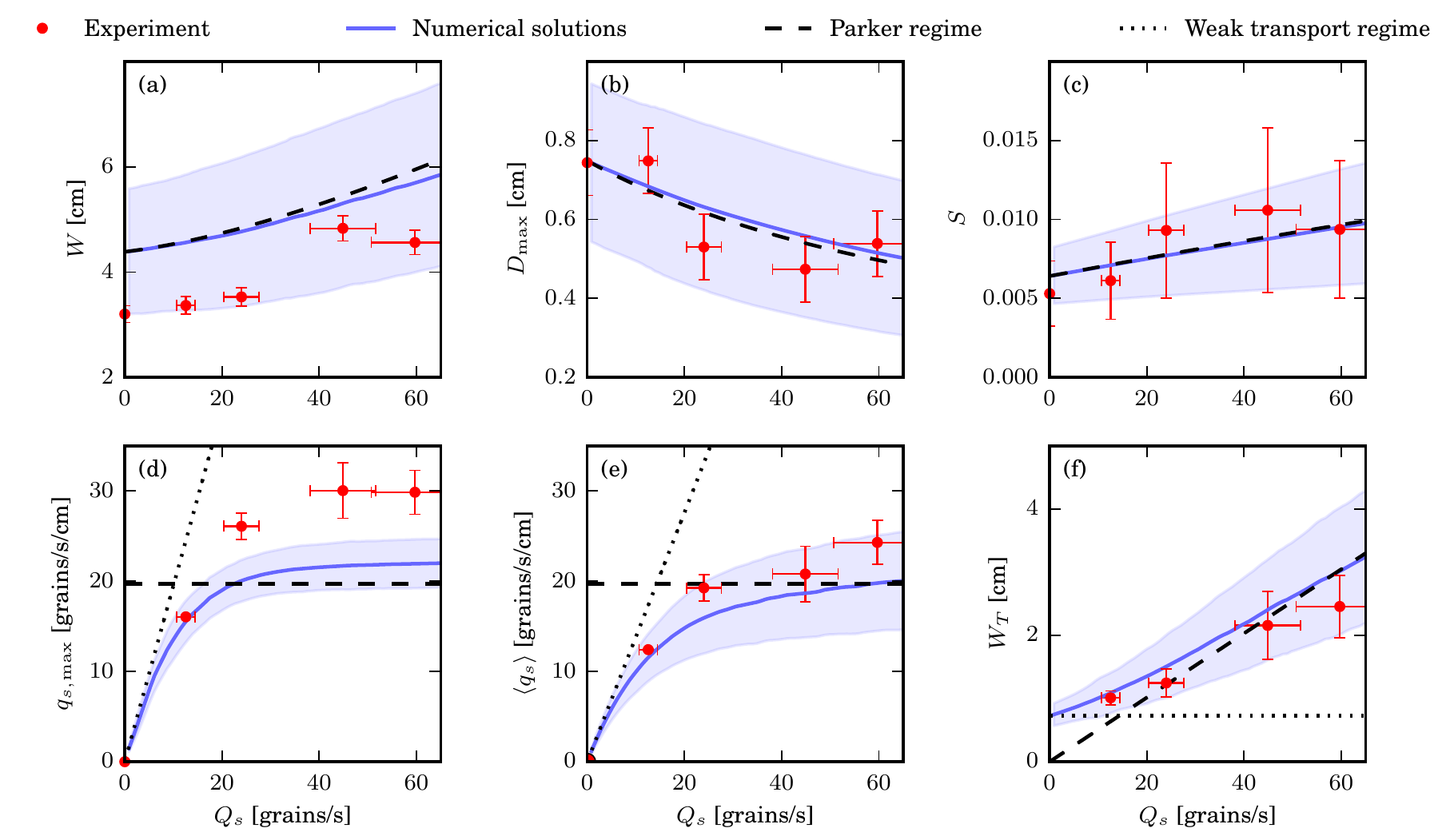}
\caption{Comparison of various river properties in dimensional units between our model and experiments of \citet{abramian2020laboratory}. Slope in panel (c) is estimated using Eq. \ref{slope_estimate}. We only show the weak transport regime for the properties related to the sediment flux. }
\label{fig:compare_dimensional_properties}
\end{figure}

As we noted in section \textit{Dependence on Water and Sediment Discharge} of the main text, the downstream slope cannot be measured directly. However, we can estimate it indirectly. The sediment flux at the river center, $q_{s,\text{max}}$, is driven only by the fluid stress, which is about $\tau \approx \rho_f g D_\text{max} S$ in the shallow-water approximation. Therefore, dropping the contributions of momentum diffusion and gravity from Eq. 13 of the main text, we find $D_\text{max} S / L_s  \approx  \mu_t + q_{s,\text{max}}/q_\mu$, and we can estimate the slope in the experiment as 
\begin{equation} \label{slope_estimate}
S \approx \frac{L_s}{D_\text{max}} \left( \mu_t + \frac{q_{s,\text{max}} }{ q_\mu} \right)  \text{  . }
\end{equation}
This is the estimate we show in Fig. \ref{fig:compare_dimensional_properties}. This estimate is not independent from our model. We can, nevertheless, use it to check the consistency between the parameters in our model and the experiments. We estimate the error for the slope calculated in this way as a combination of the errors for the quantities that enter Eq. \ref{slope_estimate}. 

The quantities we described above are all point-measurements and, therefore, are sensitive to the roughness of the bed and measurement error. On the other hand, integral quantities such as the mean sediment flux, $\langle q_s \rangle$, and transport width, $W_T$, are robust against such errors. In this case, we expect that the uncertainty of $\langle q_s \rangle$ and $W_T$ is mostly due to their natural variability in time and space, rather than by any measurement error. We, however, do not have access to measurements that would allow us to estimate the variability, so we assume that the natural variability of $\langle q_s \rangle$ is of the same order as the variability of the sediment flux profile across the channel. Therefore, we take the uncertainty of $\langle q_s \rangle$ to be equal to the uncertainty of $q_{s,\text{max}}$ --- a conservative estimate. Finally, since $W_T = Q_s / \langle q_s \rangle$, we estimate its uncertainty as a combination of the uncertainty of the sediment discharge and that of the mean sediment flux. 

\begin{table}
\centering
\caption{Table of parameters used in the experiments and the model.}
\begin{tabular}{| c | c | c | c | }
\hline
\makecell{\bf{Definition}}  &   \makecell{\bf{Notation}} &   \makecell{\bf{Value}}&  \makecell{\bf{Unit}}  \\
\hline
\makecell{Grain diameter}  &   \makecell{$d_s$} &   \makecell{$0.83 \pm 0.2$}&  \makecell{mm}  \\
\hline
\makecell{Fluid viscosity}  &   \makecell{$\nu$} &   \makecell{$10^{-5}$}&  \makecell{m$^2$s$^{-1}$}  \\
\hline
\makecell{Fluid density}  &   \makecell{$\rho_f$} &   \makecell{ $1160 \pm 5$ }&  \makecell{kg m$^{-3}$}  \\
\hline
\makecell{Sediment density}  &   \makecell{$\rho_s$} &   \makecell{$1490$} &  \makecell{kg m$^{-3}$}  \\
\hline
\makecell{Fluid discharge}  &   \makecell{$Q_w$} &   \makecell{$0.97 \pm 0.05$}&  \makecell{l min$^{-1}$}  \\
\hline
\makecell{Sediment discharge}  &   \makecell{$Q_s$} &   \makecell{$(0, 60)$}&  \makecell{grains $\text{s}^{-1}$}  \\
\hline
\makecell{Threshold Shields parameter}  &   \makecell{$\theta_t$} &   \makecell{$0.167 \pm 0.003$}&  \makecell{ None }  \\
\hline
\makecell{Friction coefficient}  &   \makecell{$\mu_t$} &   \makecell{ $0.9 \pm 0.2$}&  \makecell{None}  \\
\hline
\makecell{Sediment diffusion length}  &   \makecell{$\lambda$} &   \makecell{$0.10 \pm 0.03$}&  \makecell{mm}  \\
\hline
\makecell{Sediment flux scale}  &   \makecell{$q_\mu$ } &   \makecell{$107 \pm 30$}&  \makecell{grains cm$^{-1}$s$^{-1}$}  \\
\hline
\end{tabular}
\label{tab:1}
\end{table}

\begin{table}
\centering
\caption{River properties for each of the experiments performed by \citet{abramian2020laboratory}.}

\resizebox{0.75\columnwidth}{!}{

\begin{tabular}{| c | c | c | c |}

\cline{1-3}
\makecell{\bf{Experiment label}}  &   \makecell{\bf{Sediment discharge}} & \makecell{\bf{Unit}}\\
\cline{1-3}
\multicolumn{1}{| c |}{$1$}  &   \multicolumn{1}{ c |}{$0$} &  \multirow{5}{*}{$\text{grains s}^{-1}$} \\ 
\cline{1-2}
\multicolumn{1}{| c |}{$2$}  &   \multicolumn{1}{ c |}{$12.6 \pm 1.2$} &\\
\cline{1-2}
\multicolumn{1}{| c |}{$3$}  &   \multicolumn{1}{ c |}{$24 \pm 2.4$} &\\
\cline{1-2}
\multicolumn{1}{| c |}{$4$}  &   \multicolumn{1}{ c |}{$44.9 \pm 4.5$} & \\
\cline{1-2}
\multicolumn{1}{| c |}{$5$}  &   \multicolumn{1}{ c |}{$59.7 \pm 6.0$} & \\
\cline{1-3}
\multicolumn{1}{ c }{} \\
\multicolumn{1}{ c }{} \\
\hline
\makecell{\bf{Quantity}}  &   \makecell{\bf{Experiment label}} &   \makecell{\bf{Value}}  &   \makecell{\bf{Unit}}   \\
\hline \hline
\multirow{5}{*}{Width, $W$} & \multicolumn{1}{ c |}{$1$} & \multicolumn{1}{c |}{ $3.20 \pm 0.16$ } &  \multirow{5}{*}{$\text{cm}$} \\ \cline{2-3}
                                 & \multicolumn{1}{ c |}{$2$} & \multicolumn{1}{c |}{ $3.36 \pm 0.17$ } & \\ \cline{2-3}
                                 & \multicolumn{1}{ c |}{$3$} & \multicolumn{1}{c |}{ $3.53 \pm 0.18$ } & \\ \cline{2-3}
                                 & \multicolumn{1}{ c |}{$4$} & \multicolumn{1}{c |}{ $4.83 \pm 0.24$ } & \\ \cline{2-3}
                                 & \multicolumn{1}{ c |}{$5$} & \multicolumn{1}{c |}{ $4.56 \pm 0.23$ } & \\ \hline \hline
                                 
\multirow{5}{*}{Depth, $D_\text{max}$} & \multicolumn{1}{ c |}{$1$} & \multicolumn{1}{c |}{ $0.744 \pm 0.083$ } &  \multirow{5}{*}{$\text{cm}$} \\ \cline{2-3}
                                 & \multicolumn{1}{ c |}{$2$} & \multicolumn{1}{c |}{ $0.748 \pm 0.083$ } & \\ \cline{2-3}
                                 & \multicolumn{1}{ c |}{$3$} & \multicolumn{1}{c |}{ $0.530 \pm 0.083$ } & \\ \cline{2-3}
                                 & \multicolumn{1}{ c |}{$4$} & \multicolumn{1}{c |}{ $0.474 \pm 0.083$ } & \\ \cline{2-3}
                                 & \multicolumn{1}{ c |}{$5$} & \multicolumn{1}{c |}{ $0.539 \pm 0.083$ } & \\ \hline \hline
                                 
\multirow{5}{*}{Maximum sediment flux, $q_{s,\text{max}}$} & \multicolumn{1}{ c |}{$1$} & \multicolumn{1}{c |}{ $ 0 $ } &  \multirow{5}{*}{$\text{grains s}^{-1}\text{cm}^{-1}$} \\ \cline{2-3}
                                 & \multicolumn{1}{ c |}{$2$} & \multicolumn{1}{c |}{ $15.96 \pm 0.09$ } & \\ \cline{2-3}
                                 & \multicolumn{1}{ c |}{$3$} & \multicolumn{1}{c |}{ $24.65 \pm 1.44$ } & \\ \cline{2-3}
                                 & \multicolumn{1}{ c |}{$4$} & \multicolumn{1}{c |}{ $26.97 \pm 3.07$ } & \\ \cline{2-3}
                                 & \multicolumn{1}{ c |}{$5$} & \multicolumn{1}{c |}{ $27.42 \pm 2.44$ } & \\ \hline \hline
                                 
\multirow{5}{*}{Transport width, $W_T$} & \multicolumn{1}{ c |}{$1$} & \multicolumn{1}{c |}{ n/a } &  \multirow{5}{*}{$\text{cm}$} \\ \cline{2-3}
                                 & \multicolumn{1}{ c |}{$2$} & \multicolumn{1}{c |}{ $1.01 \pm 0.11$ } & \\ \cline{2-3}
                                 & \multicolumn{1}{ c |}{$3$} & \multicolumn{1}{c |}{ $1.24 \pm 0.22$ } & \\ \cline{2-3}
                                 & \multicolumn{1}{ c |}{$4$} & \multicolumn{1}{c |}{ $2.16 \pm 0.53$ } & \\ \cline{2-3}
                                 & \multicolumn{1}{ c |}{$5$} & \multicolumn{1}{c |}{ $2.46 \pm 0.49$ } & \\ \hline \hline

\multirow{5}{*}{Downstream slope, $S$} & \multicolumn{1}{ c |}{$1$} & \multicolumn{1}{c |}{ $0.005 \pm 0.002$ } &  \multirow{5}{*}{None} \\ \cline{2-3}
                                 & \multicolumn{1}{ c |}{$2$} & \multicolumn{1}{c |}{ $0.006 \pm 0.002$ } & \\ \cline{2-3}
                                 & \multicolumn{1}{ c |}{$3$} & \multicolumn{1}{c |}{ $0.009 \pm 0.004$ } & \\ \cline{2-3}
                                 & \multicolumn{1}{ c |}{$4$} & \multicolumn{1}{c |}{ $0.011 \pm 0.005$ } & \\ \cline{2-3}
                                 & \multicolumn{1}{ c |}{$5$} & \multicolumn{1}{c |}{ $0.009 \pm 0.004$ } & \\ \hline                                 

\end{tabular}
}
\label{tab:2}
\end{table}

\newpage

\section{Fluid induced stress, $\tau$}

In this section we will first derive Eq. 3 of the main text for the fluid stress, $\tau$. Then, we will show that this stress can be written as a series expansion that assumes the aspect ratio of the channel is large. We will show that our model, Eq. 13 of the main text, is the first term in this expansion. Finally, we will test our approximation on several tractable examples. 

\subsection{Deriving the fluid stress equation}

In this section, following \citet{devauchelle2021viscous}, we derive equation 3 of the main text for the fluid stress, $\tau$. To that end, we first integrate the Stokes' flow equation, Eq. 2 of the main text, over the vertical to get
\begin{linenomath*}
\begin{equation} \label{Integrated_Stokes_law}
\eta_v \int_{-D}^0 \frac{\partial^2 u}{\partial y^2} \text{d} z - \tau_z   = -  \rho_f g S D \text{  , }
\end{equation}
\end{linenomath*}
where the $z$-component of the stress is $\tau_z \equiv -\eta_v \left. \frac{\partial u}{\partial z} \right\vert_{z = -D}$, and we have used the boundary condition that the stress vanishes at the surface, $\tau_z(z = 0) = 0$. To get Eq. 3 of the main text, we need to pull the $y$-derivatives in Eq. \ref{Integrated_Stokes_law} outside the integral. Doing this, we get\begin{linenomath*}
\begin{equation} \label{derivative_out_2}
\int_{-D}^0 \frac{\partial^2 u}{\partial y^2} \text{d} z  = \frac{\partial^2 }{\partial y^2}\int_{-D}^0 u \text{d} z  + \frac{\text{d} D}{\text{d} y}  \left.\frac{\partial u}{\partial y} \right\vert_{z = -D}\text{  , }
\end{equation}
\end{linenomath*}
where we have used the boundary condition that the velocity vanishes at the boundary, $u(z = -D) = 0$. We can use this same condition to relate the term $\left.\frac{\partial u}{\partial y} \right\vert_{z = -D}$ to the shear stress, $\tau_z$. Namely, differentiating the boundary condition, $u(z = -D) = 0$, yields 
\begin{linenomath*}
\begin{equation} \label{du_dy_1}
 \left.\frac{\partial u}{\partial y} \right\vert_{z = -D} = \frac{\text{d} D}{\text{d} y} \left.\frac{\partial u}{\partial z} \right\vert_{z = -D}  \text{  . }
\end{equation}
\end{linenomath*}
Substituting relations Eqs. \ref{derivative_out_2} and \ref{du_dy_1} into Eq. \ref{Integrated_Stokes_law}, we find
\begin{linenomath*}
\begin{equation} \label{tau_z_equation}
\eta_v \frac{\text{d}^2 }{\text{d} y^2} (D \bar{u}) - \tau_z \left(1 + \left( \frac{\text{d}D}{\text{d} y}\right)^2 \right)   = -  \rho_f g S D \text{  , }
\end{equation}
\end{linenomath*}
where we have introduced $\bar{u}(y) \equiv \frac{1}{D}\int_{-D}^0 u \text{d} z$. Equation \ref{du_dy_1} relates the $y$ and $z$ components of the stress as $\tau_y = D' \tau_z$. This means that the total stress, $\tau$, is related to $\tau_z$ as 
\begin{linenomath*}
\begin{equation} \label{tau_vs_tau_z}
\tau = \left(\tau_z^2 + \tau_y^2\right)^{1/2} = \tau_z \left( 1 + D'^2\right)^{1/2}{  , }
\end{equation}
\end{linenomath*}
where prime stands for $\text{d}/\text{d}y$. Substituting Eq. \ref{tau_vs_tau_z} into Eq. \ref{tau_z_equation}, we can express $\tau$ as
\begin{linenomath*}
\begin{equation} \label{tau_1}
\tau = \frac{\eta_v (D \bar{u})'' + \rho_f g S D}{\left(1 + D'^2 \right)^{1/2}}  \text{  , }
\end{equation}
\end{linenomath*}
Identifying $\left(1 + D'^2 \right)^{-1/2}$ with $\cos{\phi}$, we finally retrieve Eq. 3 of the main text
\begin{linenomath*}
\begin{equation} \label{tau_final}
\tau = \left( \eta_v (D \bar{u})'' + \rho_f g S D\right) \cos{\phi} \text{  , }
\end{equation}
\end{linenomath*}
So far, we made no approximation to get here from the original Stokes equation (Eq. 2 of the main text). 



\subsection{The stress approximation as a series expansion}

In the main text, we used the shallow-water velocity to approximate the fluid stress (Eqs. 12 and 13 of the main text). In this section, we will show that this approximation is the first term in a series expansion that assumes the aspect ratio of the river is large. In this way, this approximation may be systematically improved, assuming that the series converges. 

We begin with the Stokes equation
\begin{linenomath*}
\begin{equation}
\frac{\partial^2 u}{\partial y^2} +\frac{\partial^2 u}{\partial z^2} = \frac{-gS}{\nu}
\end{equation}
\end{linenomath*}
If the channel depth, $D(y)$, varies smoothly, then the second $y$-derivative of velocity scales as $U/W^2$ while the second $z$-derivative scales as $U/D_\text{max}^2$, where $U$ is the velocity scale. Therefore, if the aspect ratio, $W/D_\text{max}$, is large, the $z$-derivative is much larger than the $y$-derivative. We can make this obvious by rescaling the variables in the Stokes equation as 
\begin{linenomath*}
\begin{equation}
\hat{{y}} = \frac{y}{W} \quad \text{ , } \quad \hat{{z}} = \frac{z}{D_\text{max}} \quad \text{ , } \quad \hat{{u}} \equiv u \frac{ \nu}{g S D_\text{max}^2}
\end{equation}
\end{linenomath*}
In these new coordinates, the channel has unit width and depth, and the fluid is driven with a unit forcing. The Stokes equation then becomes
\begin{linenomath*}
\begin{equation}
\frac{D_\text{max}^2}{W^2}\frac{\partial^2  \hat{{u}} }{\partial \hat{{y}}^2} +\frac{\partial^2 \hat{{u}}}{\partial  \hat{{z}} ^2} = -1
\end{equation}
\end{linenomath*}
We can then expand the velocity as 
\begin{linenomath*}
\begin{equation}
\hat{{u}} = \hat{{u}}^{(0)} + \hat{{u}}^{(1)} + \hat{{u}}^{(2)} + \text{...  ,}
\end{equation}
\end{linenomath*}
where $\hat{{u}}^{(0)}$ is the term of the order unity, $\hat{{u}}^{(1)}$ is the term of the order $D_\text{max}^2 / W^2$, $\hat{{u}}^{(2)}$ is the term of the order $D_\text{max}^4 / W^4$, etc. Equating orders in the Stokes equation, we get (back in the physical coordinates)
\begin{linenomath*}
\begin{align}
&\frac{\partial^2 u^{(0)}}{\partial z^2} = \frac{-gS}{\nu} \text{  , } \label{u0_partial_SI}\\
&\frac{\partial^2 u^{(0)}}{\partial y^2} + \frac{\partial^2 u^{(1)}}{\partial z^2} = 0  \text{  , ...}  \label{u1_partial_SI}
\end{align}
\end{linenomath*}
The boundary conditions for each term are the same --- velocity at each order must vanish on the channel bed and its $z$-derivative must vanish on the fluid surface. From Eq. \ref{u0_partial_SI}, we can see that the zeroth order term is in fact equal to the shallow-water velocity, $u^{(0)} \equiv u_\text{sw}$. All of the terms in the expansion are polynomial in $z$, and we can easily find each term of order $(n+1)$ based on the previous term of order $(n)$ by performing integration over $z$. In this way, we find for the first two terms:   
\begin{linenomath*}
\begin{align}
&u^{(0)} = \frac{gS}{2\nu} \left(D^2 - z^2\right) \text{  , }\\
&u^{(1)} =  \frac{gS}{4\nu} \left(D^2 - z^2\right) (D^2)'' \text{  . }
\end{align}
\end{linenomath*}
The term $u^{(1)}$ is inversely proportional to the square of the aspect ratio since the correcting factor, $(D^2)''$, is of order $D_\text{max}^2 / W^2$. From here, we can find the vertically averaged velocity as an expansion:
\begin{linenomath*}
\begin{align}
& \bar{u} = \bar{u}^{(0)} + \bar{u}^{(1)} + \text{...} \label{u_bar_expansion_SI}\\
&\bar{u}^{(0)} = \frac{g S D^2}{3\nu} \text{  , }\label{u0_bar_SI}\\
&\bar{u}^{(1)} =  \frac{gS D^2 }{6\nu} (D^2)'' \text{  . } \label{u1_bar_SI}
\end{align}
\end{linenomath*}
Equation \ref{u0_bar_SI} for the zeroth order velocity is the same as Eq. 12 of the main text for the shallow-water velocity. With this, we can find the fluid stress. Recall the exact equation for the stress, Eq. 3 of the main text,
\begin{linenomath*}
\begin{equation}\label{tau_exact_series}
\tau = \left( \rho_f g S D + \rho_f \nu (\bar{u} D)'' \right) \cos \phi \text{  . }
\end{equation}
\end{linenomath*}
We can also write the stress as an expansion 
\begin{linenomath*}
\begin{equation}\label{tau_orders} 
\tau = \tau^{(0)} + \tau^{(1)} + \text{...  ,}
\end{equation}
\end{linenomath*}
where $\tau^{(0)}$ is the leading order term, and $\tau^{(1)}$ is proportional to $D_\text{max}^2 / W^2$. Using the expansion for $\bar{u}$, we see that 
\begin{linenomath*}
\begin{align}
& \tau^{(0)} =  \rho_f g S D  \cos \phi \label{tau_orders_1} \\
& \tau^{(1)} =  \rho_f \nu (\bar{u}^{(0)} D)'' \cos \phi  = \left[\rho_f g S \frac{1}{3}\left(D^3\right)''\right] \cos \phi \text{  . } \label{tau_orders_2} 
\end{align}
\end{linenomath*}
Since $(D^3)'' \sim D_\text{max}^3/W^2$, the first-order term is about $D_\text{max}^2/W^2$ times smaller than the leading order term: $\tau^{(1)} \sim \tau^{(0)} D_\text{max}^2/W^2$. The cosine term in the above equations is $\cos \phi = \left(1 + D'^2 \right)^{-1/2}$, and, since $D'^2 \sim D_\text{max}^2 / W^2$, this term could also be expanded in a series but, for convenience, we keep the entire term here. 

The zeroth order term, $\tau^{(0)}$, with $\cos \phi \approx 1$, is the shallow-water stress that was used previously to estimate the shape of inert rivers \cite{seizilles2013width}. Our approximation (Eq. 13 of the main text) amounts to truncating the $\tau$-series at first order, $\tau \approx \tau^{(0)} + \tau^{(1)}$. In the main text, we showed that the river is formed by the combined action of fluid stress and gravity. The gravity term in the force ratio, $(D')^2$, is of the order of $D_\text{max}^2 / W^2$, and is, therefore, of the same order of magnitude as $\tau^{(1)}$. For this reason, keeping only the shallow-water term in the force ratio, $\mu$, is inconsistent, since it neglects a stress term that is comparable to gravity.

Although the expansion above may be corrected to arbitrary order, it is always local --- the stress is always a function of the depth and its derivatives. To get non-local effects, whereby the stress depends on the entire bed shape, we would have to add infinitely many terms in the expansion.

\subsection{Testing the approximation}\label{sec:SI_testing_the_approximation}

In the previous section we showed that the fluid velocity and stress can be found using a series expansion when the flow is shallow enough. In this section, we will explicitly test this expansion on several examples for which exact solutions exist.

\begin{figure}
\centering
\includegraphics[width=1.\textwidth]{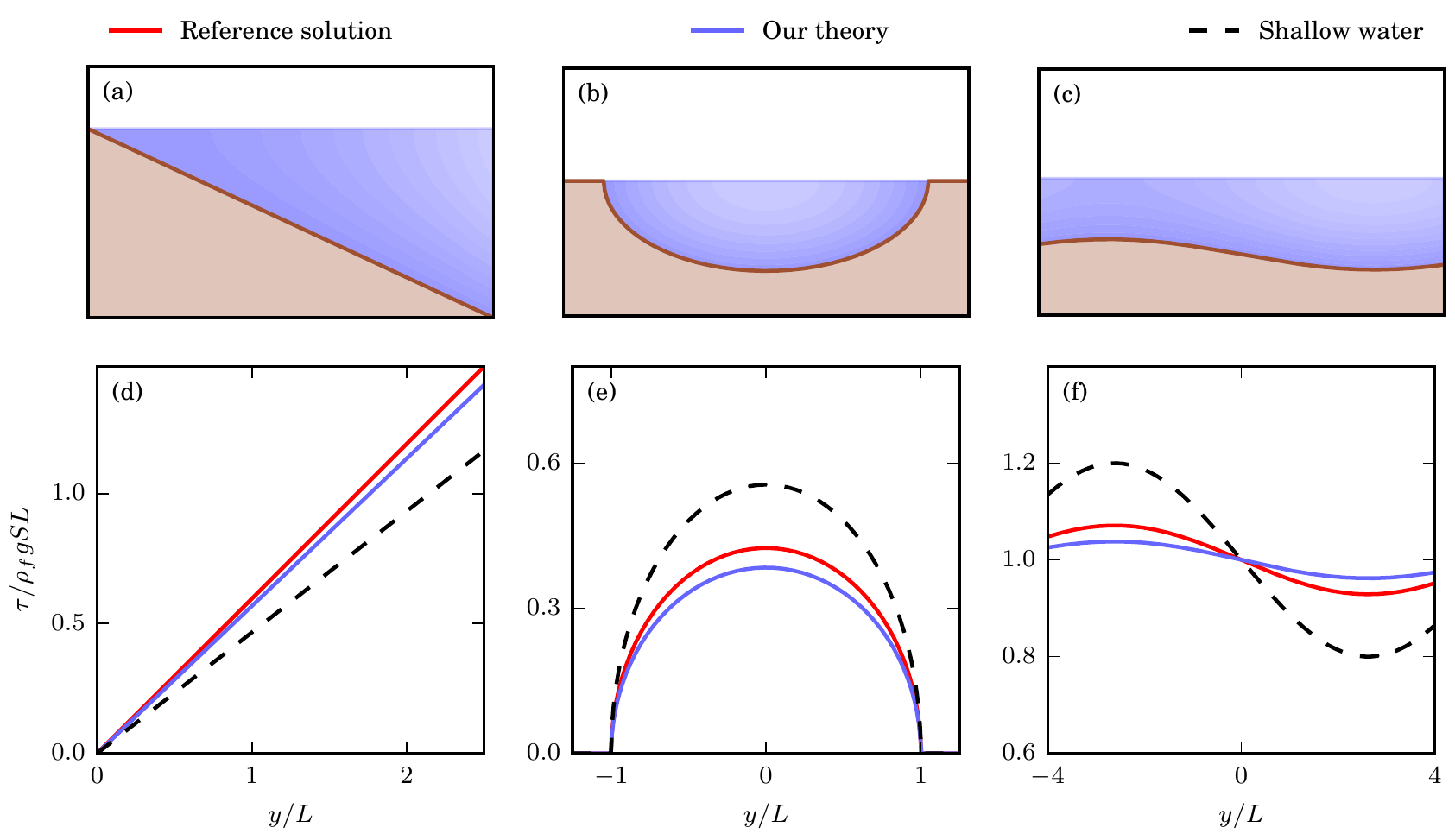}
\caption{Comparison of stress between our first order approximation (blue lines), shallow-water approximation (black dashed lines), and a analytical reference solution (red lines) in examples for which an analytical solution is available. The upper row shows the channel shape, along with the velocity field of the analytical solution (lighter blue stands for faster moving fluid). The lower row shows the normalized stress profile, $\tau(y/L) / \rho_f g S L$, where $L$ is a reference length scale that is different for each example. The spatial coordinates, $y$ and $z$, in upper and lower panels are also normalized by $L$. (a) and (d) Flow over an inclined surface of an opening angle $\phi = 25^\circ$ (Eqs. \ref{wedge_SI} to \ref{wedge_approx_tau_z}). The reference length scale in this case can be arbitrarily chosen. (b) and (e) Flow in an elliptic channel with a ratio of the semi-major to semi-minor axis $R = 1.8$ (Eqs. \ref{ellipse_SI} to \ref{tau_z_first_order_ellipse_SI}). The reference length scale is the channel width, $L = W$. (c) and (f) Flow over a surface with a small-amplitude, large-wavelength perturbation, $\delta / D_0 = 0.2$ and $kD_0 = 0.9$ (Eqs. \ref{perturbation_SI} to \ref{tau_z_first_order_perturbation_SI}). The reference length scale is the unperturbed channel depth, $L = D_0$. }
\label{fig:corrections_to_shallow_water}
\end{figure}

As a first example, we look at the flow over an inclined surface with an opening angle $\phi$, that we assume to be small (Fig. \ref{fig:corrections_to_shallow_water}a). In this case, the depth profile is given by 
\begin{linenomath*}
\begin{equation} \label{wedge_SI}
D(y) = y \tan \phi \text{  . } 
\end{equation}
\end{linenomath*}
This problem is not well-posed since we do not specify the boundary condition on the open edge of the wedge. Nevertheless, if the conditions far away do not matter, the solution of Stokes equation and the corresponding stress should be 
\begin{linenomath*}
\begin{align}
& u = \frac{ g S }{2 \nu} \frac{ D^2 - z^2}{1- (\tan \phi)^2} \text{  , } \label{wedge_exact_u}\\
& \tau_z = \frac{gS\rho_f D }{1- (\tan \phi)^2} \text{  , } \label{wedge_exact_tau_z}
\end{align}
\end{linenomath*}
where $\tau_z$ is the vertical component of the stress. We can see that the exact solution has the same form as the shallow-water approximation with the correction factor $1- (\tan \phi)^2$. Our approximation to first order yields  
\begin{linenomath*}
\begin{align}
& u = \frac{ g S }{2 \nu} \left( D^2 - z^2 \right) \left(1 + (\tan \phi)^2\right) \text{  , } \label{wedge_approx_u}\\
& \tau_z = g S \rho_f D \left( 1+ (\tan \phi)^2 \right) \text{  , } \label{wedge_approx_tau_z}
\end{align}
\end{linenomath*}
where we have expanded the cosine terms in Eq. \ref{tau_orders_1} and \ref{tau_orders_2} to first order in $\phi$. Equations \ref{wedge_approx_u} and \ref{wedge_approx_tau_z} show that our approximation is correct to first order in $\phi$. We compare the stress in our approximation to the exact solution, and to the shallow-water stress in Fig. \ref{fig:corrections_to_shallow_water}d. For $\tan \phi \geq 1$, our approximation breaks down. In fact, the exact solution given by Eqs. \ref{wedge_exact_u} and \ref{wedge_approx_tau_z} also breaks down at $\tan \phi = 1$. For large angles, the exact solution we presented above is not viable because, in that case, the boundary condition at the far end cannot be ignored. This is an example of a non-local effect we mentioned in the previous section --- for $\tan \phi \geq 1$, our approximation breaks down because it can only relate the flow to the local bed shape. 

As our second example, we consider the flow in an elliptic channel with a ratio of semi-major to semi-minor axes equal to $R = W / (2 D_\text{max})$, that we assume to be large (Fig. \ref{fig:corrections_to_shallow_water}b). The depth profile is given by 
\begin{linenomath*}
\begin{equation}\label{ellipse_SI}
D(y) = \frac{1}{R} \sqrt{\left( \frac{W}{2} \right)^2-y^2}\text{  , } 
\end{equation}
\end{linenomath*}
The exact solution of the Stokes equation in such a channel is
\begin{linenomath*}
\begin{align}
& u =  \frac{ g S }{2 \nu} \frac{  \left( D^2 - z^2 \right) R^2  }{ (1+ R^2) } \\
& \tau_z = \frac{ g S \rho_f D R^2 }{1 + R^2} \text{  , }
\end{align}
\end{linenomath*}
Our approximation to first order gives
\begin{linenomath*}
\begin{align}
& u = \frac{ g S }{2 \nu}  \left( D^2 - z^2 \right) \left( 1 - \frac{1}{R^2} \right) \\
& \tau_z = g S \rho_f D \left(1 -  \frac{1}{R^2} \right)  \label{tau_z_first_order_ellipse_SI}
\end{align}
\end{linenomath*}
Our approximation provides the correct first order when $R$ is large, but fails when $R \geq 1$. The reason is the same as before --- in a narrow channel, rather than being controlled by the local depth configuration, the fluid velocity becomes dominated by the side walls. We compare our approximation with the exact solution and the shallow-water approximation in Fig. \ref{fig:corrections_to_shallow_water}e.

Finally, we consider the flow the flow over a sinusoidally perturbed bed (Fig. \ref{fig:corrections_to_shallow_water}c). The depth profile in this case is given by 
\begin{linenomath*}
\begin{equation}\label{perturbation_SI}
D(y) = D_0 + \delta \sin(ky) \text{  , } 
\end{equation}
\end{linenomath*}
where $\delta$ is the amplitude and $k$ is the wavenumber of the perturbation. If $\delta$ is small, the full Stokes equation can be linearized to find the stress. \citet{abramian2019streamwise} derived the expression for the stress in this case:
\begin{linenomath*}
\begin{align} \label{linear_perturbation}
\tau_z = g S \rho_f  D_0 + g S \rho_f \delta \left( 1 - k D_0 \text{tanh}(k D_0)  \right) \sin(ky)  \text{  . }
\end{align}
\end{linenomath*}
We expect our approximation to work when the wavelength of the perturbation is large compared with the flow depth (i.e. when $k D_0$ is small), but should not be limited by $\delta$. Therefore, we can compare our approximation to the result of \citet{abramian2019streamwise} when both $\delta / D_0$ and $k D_0$ are small. In this case, our approximation yields:
\begin{linenomath*}
\begin{align}
\tau_z = g S \rho_f D_0 + g S \rho_f \delta \left( 1 - (k D_0)^2  \right) \sin(ky) \label{tau_z_first_order_perturbation_SI}
\end{align}
\end{linenomath*}
Again, this is the correct first order expansion of Eq. \ref{linear_perturbation} for small $k D_0$. In the classical shallow-water theory, the stress, $\tau_z = g S \rho_f D$, is always in phase with the depth perturbation. Conversely, in the linearization of \citet{abramian2019streamwise}, Eq. \ref{linear_perturbation}, the phase of the stress can reverse when the wavelength of the perturbation is small enough, so that the stress maximum is where the flow is shallowest. This feature of the full linearized equation is reproduced in our approximation. This shows that our approximation can capture qualitative effects of the cross-stream diffusion of momentum. We compare our first order approximation, linearization of \citet{abramian2019streamwise} (Eq. \ref{linear_perturbation}), and the shallow-water approximation for a sinusoidal perturbation in Fig. \ref{fig:corrections_to_shallow_water}f.

It is clear that our approximation cannot always work. For example, in a rectangular channel, our approximation predicts uniform stress and velocity above the entire flat bottom, to all orders of the approximation. This is clearly not the case in reality, since the fluid must slow down near the vertical walls. The effect of the walls is, again, an example of a non-locality --- the velocity of the fluid is not only determined by the local depth and its derivatives, but is also affected by the far away walls. Therefore, in all cases we considered above, the failure of our approximation was related to the non-locality of the velocity field. 

\newpage

\section{River profile as a function of water and sediment discharges}

As we mentioned in the main text, our model is completely determined by setting five parameters --- $\mu_t$, $\lambda$, $L_s$, $S$, and $\xi$. Of those, $\mu_t$, $\lambda$, and $L_s$ are directly measurable in the experiments, while $S$ and $\xi$ depend implicitly on discharges $Q_w$ and $Q_s$. In this section we describe how we numerically find this relation to get the river profiles as a function of the discharges of water and sediment. 

\subsection{Solving the boundary value problem}
To get the river as a function of the discharges, we first need to be able to solve our model, Eq. 13 of the main text, for given values of the parameters $\mu_t$, $\lambda$, $L_s$, $S$, and $\xi$. This presents a slight challenge. We described this model as a second order boundary value problem with boundary conditions $D = 0$ when $y = 0$, and $D' = 0$ when $y = - W/2$. There are two problems here --- first, the second derivative, $D''$,  given by Eq. 13 of the main text, diverges when $D \rightarrow 0$, and, second, the width of the river, $W$, is unknown a priori, so the second boundary condition is ill-defined. To deal with the first issue, we expand the solution near $D = 0$, so that we only consider depths larger than some small value, $\epsilon$. Thus, we change the first boundary condition to $D = \epsilon$ when $y = 0$ (we can arbitrarily shift the solution along $y$ due to translational invariance). To deal with the second issue, we replace the second boundary condition with a new condition for $D'$ that, like the first boundary condition, starts at $y = 0$. In this way, instead of dealing with a boundary value problem, we can solve a simpler initial value problem. To do this, we need to approximate the slope near the river bank, $D'$, when the depth, $D = \epsilon$, is small. We use the fact that $D''$, which is a known function of $D$ and $D'$, diverges when $D \rightarrow 0$. In particular, on the left river bank (when $D = \epsilon$ and $D' > 0$), $D'' \rightarrow \infty$ as $\epsilon \rightarrow 0$ for all solutions with $D'$ smaller than that of the river, while $D'' \rightarrow -\infty$ for all solutions with $D'$ greater than that of the river. Therefore, for small $D$, the river approximately lies on the curve $D''(D,D') = 0$. We can thus invert the relation $D''(\epsilon,D') = 0$ to find $D'$ when $y = 0$, therefore defining the second initial condition. In summary, we solve the initial value problem, $D = \epsilon$ and $D''(\epsilon,D') = 0$ when $y = 0$, by forward stepping and we stop when we reach the river center, $D' = 0$. Once we find the left bank, the right bank follows by symmetry. 

\subsection{Interpolating the model}
\begin{figure}
\centering
\includegraphics[width=1.\textwidth]{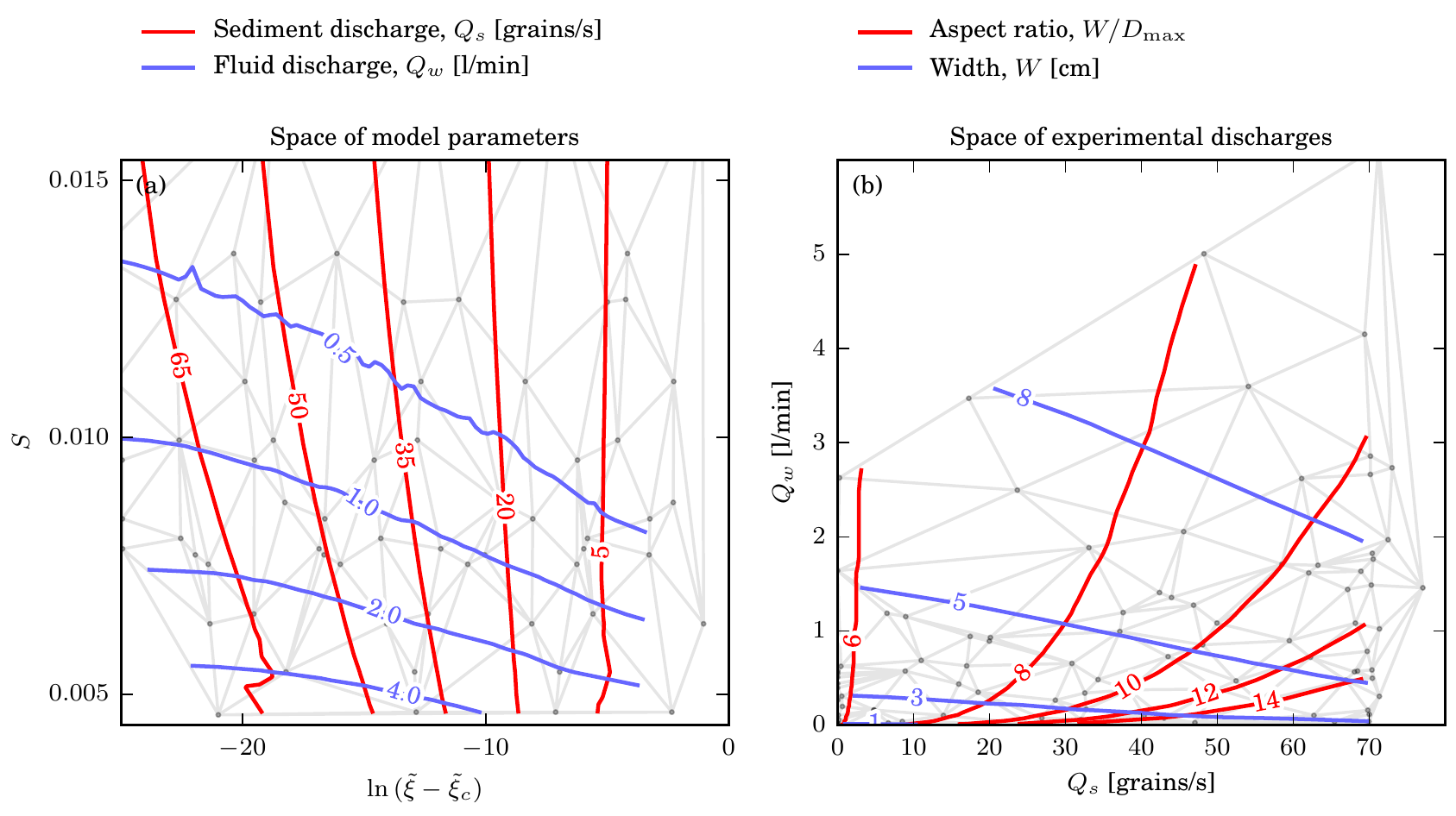}
\caption{An interpolation grid of solutions used to find model dependence on discharges of fluid and sediment. Each grey point is a solution to our model with different values of $S$ and $\xi$ (and fixed $\mu_t$, $\lambda$, $L_s$). For visual clarity, we only show every fifth point. (a) Interpolation grid in the space of model parameters, $\xi$ and $S$. The horizontal axis shows the logarithm of the difference between $\tilde{\xi}$ and the limiting value $\tilde{\xi}_c$ which is approximately proportional to the sediment discharge. Solid lines correspond to constant sediment discharge (red lines; numbers denote the value in $\text{grains s}^{-1}$) and fluid discharge (blue lines; numbers denote the value in $\text{l min}^{-1}$) found after interpolation. The non-smooth appearance of some of these lines for small and large values of $S$ is due to the sparseness of the grid. (b) Interpolation grid in the space of fluid and sediment discharge, $Q_w$ and $Q_s$. Each point on this grid has a corresponding point in panel (a). Solid lines correspond to constant aspect ratio (red lines) and width (blue lines; numbers denote the value in $\text{cm}$). }
\label{fig:interpolation_grid}
\end{figure}

Once we can find a river for given model parameters, we fix $\mu_t$, $\lambda$, and $L_s$ to their experimental values and then solve our model (Eq. 13 of the main text) for multiple values of $S$ and $\xi$ to create a grid of solutions for a range of model parameters (Fig. \ref{fig:interpolation_grid}a). For each numerical solution in this grid, we can find $Q_w$ and $Q_s$ using Eq. 15 of the main text. Interpolating over this grid, we then find $S$ and $\xi$ (or any other property of the river rivers such as the width or depth) as functions of $Q_w$ and $Q_s$. 

One problem is how to choose the range of $S$ and $\xi$ for the solution grid. Namely, increases in the sediment discharge are controlled by minute differences of $\tilde{\xi}$ (on the order of $10^{-9}$) from $\tilde{\xi}_c$, where $\tilde{\xi} \equiv \xi S / L_s$ (see section \textit{Inert, active, and limiting river} of the main text). To probe a significant range of sediment discharge, we need to explore $\tilde{\xi}$ in a very narrow range around $\tilde{\xi}_c$. We do not know the value of $\tilde{\xi}_c$ a priori, and we have to find it numerically. Moreover, the value of $\tilde{\xi}_c$ changes with $S$, so for each $S$, we need to independently estimate $\tilde{\xi}_c$ to high precision. Since there are no river-like solutions below $\tilde{\xi}_c$, we can easily identify whether a solution with given parameters $S$ and $\xi$ has $\tilde{\xi} > \tilde{\xi}_c$ or $\tilde{\xi} < \tilde{\xi}_c$. Therefore, we can find $\tilde{\xi}_c$ by a shooting method. Moreover, in the process of approaching $\tilde{\xi}_c$ by interval halving, we solve our model for multiple values of $\xi$, so we can immediately use the solutions with $\tilde{\xi} < \tilde{\xi}_c$ in our interpolation grid. This also has the useful property that exponentially approaching $\tilde{\xi}_c$ leads to approximately linearly increasing $Q_s$. Therefore, by finding $\tilde{\xi}_c$, we probe the space of sediment discharge more or less uniformly. Once we find $\tilde{\xi}_c$ for a given $S$, we add the solutions to the interpolation grid, we choose randomly a different $S$, and we repeat the procedure. 

To probe the relevant range of $S$ that corresponds roughly to the fluid discharge in the experiments, $Q_w$, we look at inert rivers ($Q_s = 0$). Using Eq. 15 of the main text, we can write the fluid discharge as 
\begin{linenomath*}
\begin{equation}\label{Qw_nondim_SI}
Q_w = \frac{ g L_s^4 }{ \nu S^3} \int_{-\tilde{W}/2}^{\tilde{W}/2} \frac{\tilde{D}^3}{3} \text{d}\tilde{y} \text{  , }
\end{equation}
\end{linenomath*}
where tildes stand for quantities made non-dimensional with the length scale, $L_s/S$, as defined in Eq. 16 of the main text. The term $\tilde{Q}_w = \int \tilde{D}^3 / 3 \text{d}\tilde{y}$ is the non-dimensional fluid discharge that depends on the shape of the river, but not its size. As we have discussed in section \textit{Inert, active, and limiting river} of the main text, the shape of non-dimensional inert rivers only depends on $\mu_t$. So, for an inert river, $\tilde{Q}_{w,0}(\mu_t)$ is independent of the fluid discharge. Therefore, we find that    
\begin{linenomath*}
\begin{equation}\label{S_inert_SI}
S_0 = \left( \frac{ g L_s^4 \tilde{Q}_{w,0}(\mu_t)}{ \nu Q_w} \right)^{1/3} \quad \text{ when } \quad Q_s = 0 \text{  . }
\end{equation}
\end{linenomath*}
Based on this expression, we estimate the order of magnitude of the experimental slope. Then, taking values between, for example, $S_0/5$ and $5 S_0$, ensures that we cover the range $S$ that is relevant for our experiment, even when the sediment discharge is finite. 

\subsection{Dependence on water and sediment discharges}

\begin{figure}
\centering
\includegraphics[width=1.\textwidth]{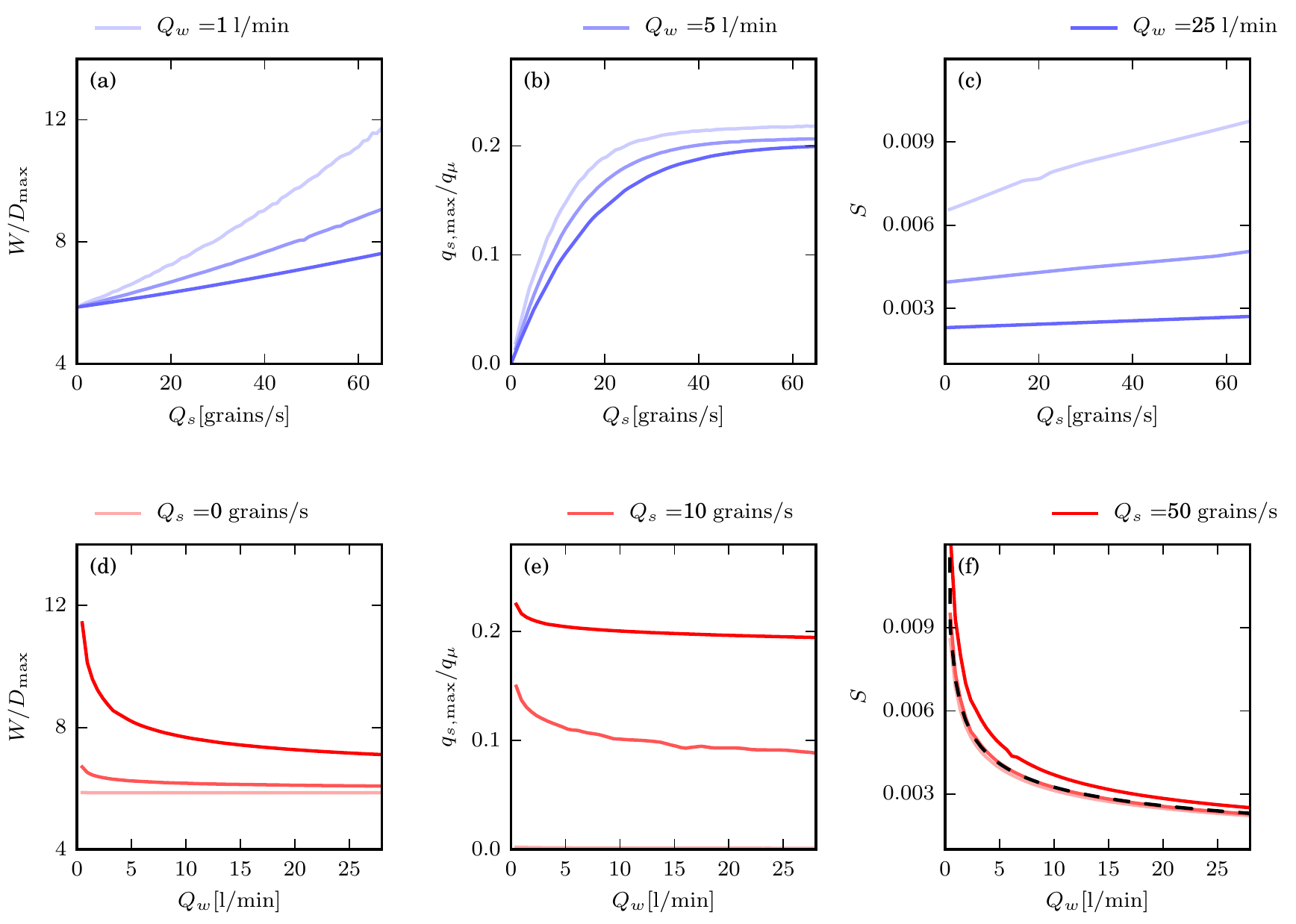}
\caption{(a-c) River properties in our model as a function of sediment discharge, $Q_s$, for various values of fluid discharge, $Q_w$. (a) Aspect ratio, $D_\text{max}/W$. (b) Maximum sediment flux, $q_{s,\text{max}} / q_\mu$. (c) Downstream slope, $S$. (d-f) River properties in our model as a function of fluid discharge, $Q_w$, for various values of sediment discharge, $Q_s$. (d) Aspect ratio, $D_\text{max}/W$. (e) Maximum sediment flux, $q_{s,\text{max}} / q_\mu$. (f) Downstream slope, $S$. The black dashed line corresponds to a curve $S = (Q_w^* / Q_w)^{1/3}$, where $Q_w^* \equiv gL_s^4  \tilde{Q}_{w,0} / \nu \approx 2.6 \times 10^{-7} \text{ l}\text{ min}^{-1}$ is the characteristic fluid discharge. }
\label{fig:plot_just_numerics_with_Qw_1}
\end{figure}

In Fig. \ref{fig:plot_just_numerics_with_Qw_1}, we show how several river properties depend on the discharges of fluid and sediment in our model. Namely, the aspect ratio increases with the sediment discharge, and only depends on the fluid discharge for large values of the sediment discharge.The shape of an inert river ($Q_s = 0$) is independent of the fluid discharge. The maximum sediment flux, $q_{s,\text{max}}$, increases with the sediment discharge and is largely independent of the fluid discharge. In fact, for large fluid discharge, $Q_w$, the maximum sediment flux saturates at $q_\mu(\tilde{D}_{\text{max},0} - \mu_t)$, as predicted by the Parker regime. Figure \ref{fig:plot_just_numerics_with_Qw_1}f shows that the downstream slope, $S$, scales approximately as $Q_w^{-1/3}$ (for an inert river, this scaling is exact, Eq. \ref{S_inert_SI}). Therefore, the scale of the river, given by $L_s/S$ roughly increases with fluid discharge as $Q_w^{1/3}$. The sediment discharge affects the slope, $S$, and the size of the river only slightly (Fig. \ref{fig:plot_just_numerics_with_Qw_1}c).

\newpage

\section{Non-dimensional river model}

In section \textit{Inert, active, and limiting river} of the main text, we discussed our model in non-dimensional form (Eq. 17 of the main text). In this section, we discuss the mathematical properties of Eq. 17, such as its fixed points and its phase portrait. These properties clarify why an infinite, limiting river exists in our model. In addition, we support claims made in sections \textit{Inert, active, and limiting river} and \textit{The Parker regime} of the main text about inert and infinite rivers. 

\subsection{Fixed points}

Since Eq. 17 of the main text is of second order, the second derivative of the depth, $\tilde{D}''$, is a function of $\tilde{D}$ and $\tilde{D}'$. Thus, we can view it as a dynamical system that can be integrated by forward-stepping, starting from a given value of $\tilde{D}$ and $\tilde{D}'$. 

Fixed points of this equation are defined as points in the $\tilde{D}$-$\tilde{D}'$ space (the phase space), that give constant solutions under integration. This is satisfied when $\tilde{D}' = 0$ and $\tilde{D}'' = 0$. Inserting these conditions into Eq. 17 of the main text yields
\begin{linenomath*}
\begin{equation}\label{fixed_point_cond_SI}
\tilde{D} = \mu_t + e^{(\tilde{D} - \tilde{\xi})/ \tilde{\lambda}} \text{  . }
\end{equation}
\end{linenomath*}
Solving this transcendental equation for $\tilde{D}$ yields the depths of the fixed points. For given values of the parameters $\mu_t$, $\tilde{\lambda}$, and $\tilde{\xi}$, Eq. \ref{fixed_point_cond_SI} has zero, one, or two solutions. We find that when $\tilde{\xi} = \tilde{\xi}_\text{bif}$, Eq. \ref{fixed_point_cond_SI} has only one solution at depth $\tilde{D}_\text{bif}$ (the subscript ``bif'' stands for ``bifurcation''), where 
\begin{linenomath*}
\begin{align}
& \tilde{\xi}_\text{bif} = \mu_t + \tilde{\lambda} (1-\ln{\tilde{\lambda}}) \text{  , } \label{xi_bif_SI} \\
& \tilde{D}_\text{bif} =  \mu_t + \tilde{\lambda} \label{D_bif_SI} \text{  . }
\end{align}
\end{linenomath*}
For $\tilde{\xi} < \tilde{\xi}_\text{bif}$, there is no fixed point, while for $\tilde{\xi} > \tilde{\xi}_\text{bif}$ there are two (at depths $\tilde{D}_1$ and $\tilde{D}_2$; brown circle and star in Figs. \ref{fig:plot_phase_diagram_evolution_and_solutions_v2_1} and \ref{fig:plot_phase_diagram_evolution_all_xi_with_bifurcation_diagram_1}). These fixed points represent flat solutions that extend to infinity in the $\tilde{y}$-direction (Fig. \ref{fig:plot_phase_diagram_evolution_and_solutions_v2_1}c). 


\subsection{Phase portrait}

We now explore how the shape of the river depends on the model parameters $\mu_t$, $\tilde{\lambda}$, and $\tilde{\xi}$ by looking at the phase portrait, i.e. the trajectories described by Eq. 17 in the $\tilde{D}$-$\tilde{D}'$ space \cite{strogatz2018nonlinear}. In addition, we use the phase portrait to unambiguously show that there exists an infinite, limiting river. 

\begin{figure}
\centering
\includegraphics[width=1\textwidth]{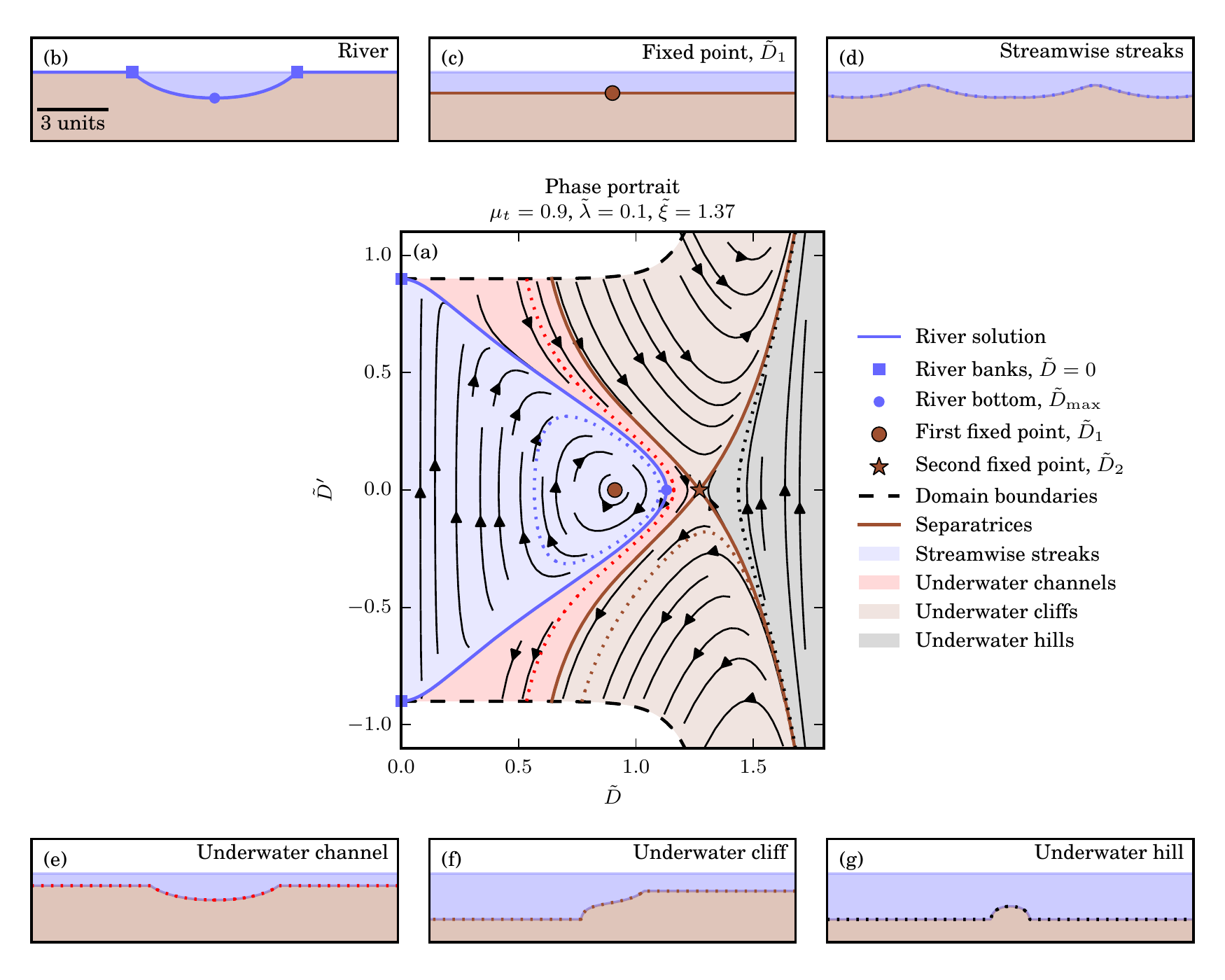}
\caption{ (a) A phase portrait of Eq. 17 of the main text for $\mu_t = 0.9$, $\tilde{\lambda} = 0.1$, and $\tilde{\xi} = 1.37$. The horizontal axis is the non-dimensional depth, $\tilde{D}$, and the vertical axis is its $\tilde{y}$-derivative, $\tilde{D}'$. Black dashed curves mark the boundaries of the region of $\tilde{D}$-$\tilde{D}'$ space on which Eq. 17 is well-defined. Black trajectories represent solutions to Eq. 17 for different initial conditions. The blue line represents the river solution shown in panel (b), while the blue dot and blue squares represent the river center and banks. The river solution is unique and acts as a separatrix between two regions of qualitatively different solutions in the phase space (streamwise streaks and underwater channels). Brown circle and brown star represent the two fixed points at depths $\tilde{D}_1$ and $\tilde{D}_2$. Brown lines are solutions ending in the second fixed point and separate regions of the phase space with qualitatively different solutions. A constant depth solution corresponding to the first fixed point, $\tilde{D}_1$, is shown in panel (c). Different colored shadings are regions of the phase space with qualitatively different solutions. An example of a solution from each of these regions is marked with a colored dotted line and shown in panels (d)-(g).}
 \label{fig:plot_phase_diagram_evolution_and_solutions_v2_1}
\end{figure}

In Fig. \ref{fig:plot_phase_diagram_evolution_and_solutions_v2_1}a, we show an example of a phase portrait for given parameters $\mu_t$, $\tilde{\lambda}$, and $\tilde{\xi}$. Each trajectory in the phase portrait corresponds to a bed profile that satisfies Eq. 17 under different boundary conditions (black lines in Fig. \ref{fig:plot_phase_diagram_evolution_and_solutions_v2_1}a). Depending on the boundary conditions, Eq. 17 has multiple, qualitatively different solutions (Fig. \ref{fig:plot_phase_diagram_evolution_and_solutions_v2_1}d-g). However, for a given set of parameters, $\mu_t$, $\tilde{\lambda}$, and $\tilde{\xi}$, there exists only one channel-like solution of Eq. 17 with two banks where the depth vanishes (blue line in Fig. \ref{fig:plot_phase_diagram_evolution_and_solutions_v2_1}a and b). Such a solution is the only one that can represent a river transporting finite amounts of water and sediment. 

\begin{figure}
\centering
\includegraphics[width=1\textwidth]{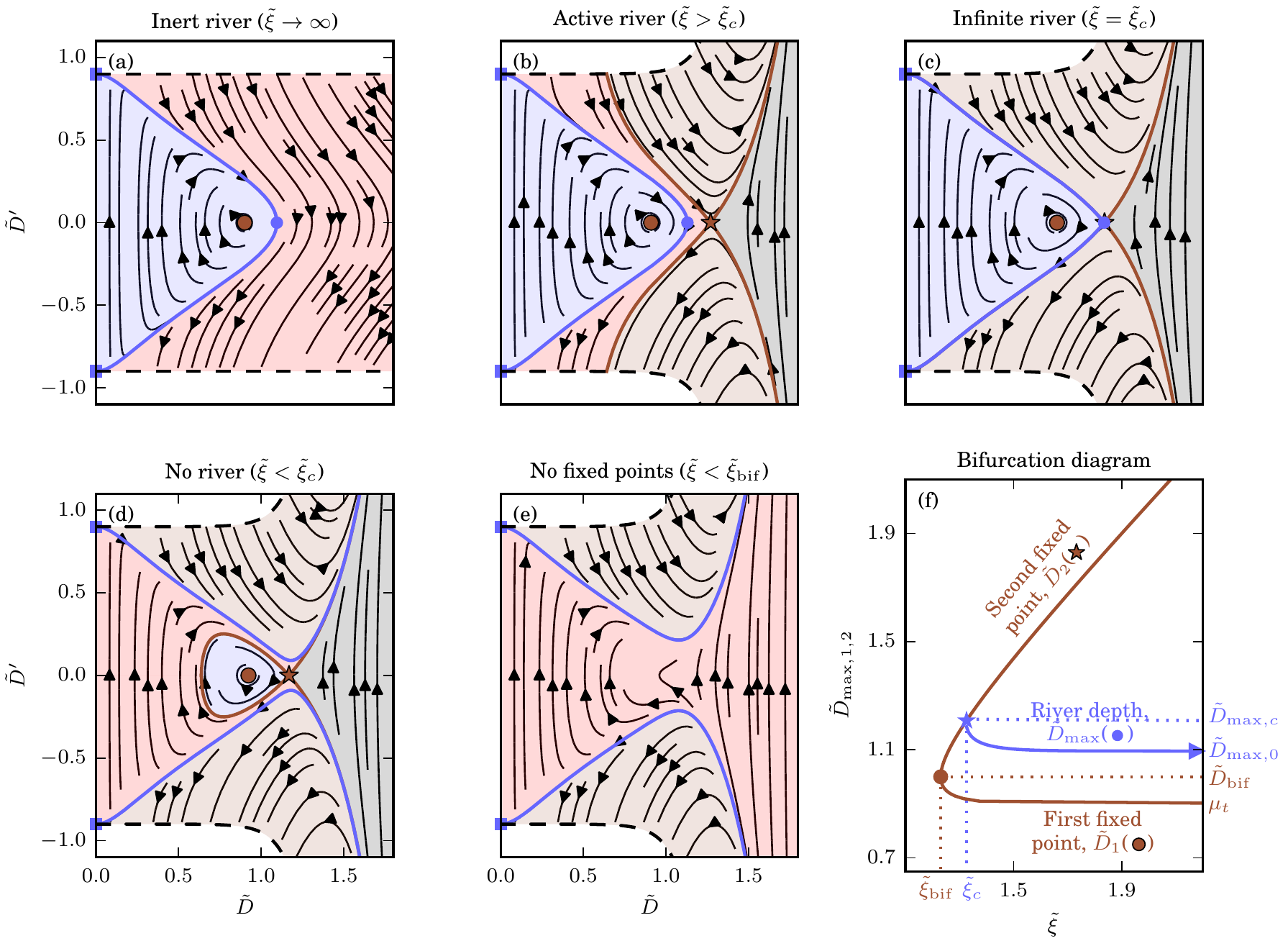}
\caption{(a)-(e) Phase portraits of Eq. 17 of the main text for $\mu_t = 0.9$, $\tilde{\lambda} = 0.1$, and varying $\tilde{\xi}$. Notation is the same as in Fig. \ref{fig:plot_phase_diagram_evolution_and_solutions_v2_1}. (a) Inert river ($\tilde{\xi} \rightarrow \infty$). There exists only one fixed point at $\tilde{D}_1 = \mu_t$. (b) Active river ($\tilde{\xi} = 1.37 > \tilde{\xi}_c$). River solution passes between the two fixed points. (c) Infinite, limiting river ($\tilde{\xi} = 1.3237 \approx \tilde{\xi}_c$). The river solution reaches a maximum depth at the second fixed point, $\tilde{D}_{\mathrm{max},c} = \tilde{D}_2$. (d) No river solution, but fixed points still exist ($\tilde{\xi} = 1.3 < \tilde{\xi}_c$).  The blue line shows the solutions starting at $\tilde{D} = 0$. These solutions do not represent a river since they do not join at the center. (e) No river or fixed point solutions ($\tilde{\xi} = 1.2 < \xi_\text{bif} \approx 1.23$). The two fixed points merge and disappear. (f) The bifurcation diagram for $\mu_t = 0.9$ and $\tilde{\lambda} = 0.1$. The brown lines represent the depths of the fixed points as a function of $\tilde{\xi}$ --- lower branch corresponds to the first fixed point, $\tilde{D}_1$ (brown circle in panels a-e), while the upper branch corresponds to the second fixed point, $\tilde{D}_2$ (brown star in panels a-e). When $\tilde{\xi}\rightarrow \infty$, the first fixed point approaches the friction coefficient ($\tilde{D}_1 \rightarrow \mu_t$), while the second tends to infinity ($\tilde{D}_2 \rightarrow \infty$). The two fixed points meet at $(\tilde{\xi}, \tilde{D}) = (\tilde{\xi}_\text{bif}, \tilde{D}_\text{bif})$ (brown dot; $\tilde{\xi}_\text{bif} \approx 1.23$, $\tilde{D}_\text{bif} = 1$). The blue line corresponds to the river depth, $\tilde{D}_\text{max}$ (blue dot in panels a-e). When $\tilde{\xi}\rightarrow \infty$, the river approaches the inert river depth, $\tilde{D}_{\text{max},0}$ (blue triangle). The river meets the second fixed point at $(\tilde{\xi}, \tilde{D}) = (\tilde{\xi}_c, \tilde{D}_{\text{max},c})$ (blue star; $\tilde{\xi}_c \approx 1.3237$, $\tilde{D}_{\text{max},c} \approx 1.21$). At this point the river is flat and infinitely wide.  }
 \label{fig:plot_phase_diagram_evolution_all_xi_with_bifurcation_diagram_1}
\end{figure}

In Fig. \ref{fig:plot_phase_diagram_evolution_all_xi_with_bifurcation_diagram_1}, we show how the phase portrait changes as we change $\tilde{\xi}$, but keep $\mu_t$ and $\tilde{\lambda}$ fixed. For $\tilde{\xi} \rightarrow \infty$ (Fig. \ref{fig:plot_phase_diagram_evolution_all_xi_with_bifurcation_diagram_1}a), the river is inert and reaches a maximum depth $\tilde{D}_{\mathrm{max},0} > \mu_t$, while there exists only one fixed point at $\tilde{D}_1 \rightarrow \mu_t$. Decreasing $\tilde{\xi}$ to finite values (Fig. \ref{fig:plot_phase_diagram_evolution_all_xi_with_bifurcation_diagram_1}b), the second fixed point appears, and the river depth lies between the two points, $\tilde{D}_1 < \tilde{D}_{\mathrm{max}} < \tilde{D}_2$. For a particular value $\tilde{\xi} = \tilde{\xi}_c$ (Fig. \ref{fig:plot_phase_diagram_evolution_all_xi_with_bifurcation_diagram_1}c), the river solution passes through the second fixed point, $\tilde{D}_{\mathrm{max},c} = \tilde{D}_2$. Therefore, the river becomes infinite in the $\tilde{y}-\tilde{z}$ space. The value of $\xi_c$ depends on $\mu_t$ and $\tilde{\lambda}$, but it exists for any value of these parameters. 

Reducing $\tilde{\xi}$ below this critical value, $\tilde{\xi}_c$, changes the phase portrait such that no river solution can exist anymore (Fig. \ref{fig:plot_phase_diagram_evolution_all_xi_with_bifurcation_diagram_1}d) --- the river solution does not exist since the two banks (solutions starting at $\tilde{D} = 0$) do not join at the center. Reducing $\tilde{\xi}$ even further, below $\tilde{\xi}_\text{bif}$ (Fig. \ref{fig:plot_phase_diagram_evolution_all_xi_with_bifurcation_diagram_1}e), the two fixed points merge and disappear so that there are neither river nor fixed point solutions. 

We show the positions of the fixed points, $\tilde{D}_1$ and $\tilde{D}_2$, and the river depth, $\tilde{D}_\text{max}$, as a function of $\tilde{\xi}$ in the bifurcation diagram, Fig. \ref{fig:plot_phase_diagram_evolution_all_xi_with_bifurcation_diagram_1}f. There, we can see that, as $\tilde{\xi}$ decreases from $\tilde{\xi} \rightarrow \infty$ to finite values, the river depth increases until it meets with $\tilde{D}_2$ at $\tilde{\xi} = \tilde{\xi}_c$. 

The existence of $\tilde{\xi}_c$ implies a limiting flux $q_{s,c}$ of the river (section \textit{Inert, active, and limiting river} of the main text). By the same reasoning, one would be tempted to conclude that the existence of $\tilde{\xi}_\text{bif}$ indicates a maximum sediment flux for a flat bed (that could be realized, for example, in a flume experiment). This is, however, not true --- when $\tilde{\xi} \rightarrow \infty$, the fixed points at $\tilde{D}_1 = \mu_t$ and $\tilde{D}_2 \rightarrow \infty$ have sediment fluxes $q_{s,1} = 0$ and $q_{s,2} \rightarrow \infty$. Decreasing $\tilde{\xi}$ to finite values, $q_{s,1}$ increases while $q_{s,2}$ decreases until they meet, thereby covering the entire range of possible sediment flux values from 0 to $\infty$. A flat bed in a flume can, therefore, carry any sediment flux, at least in principle. In practice, however, some of these solutions may become unstable to perturbations \cite{abramian2019streamwise}.

\subsection{Inert river}

\begin{figure}
\centering
\includegraphics[width=0.66\textwidth]{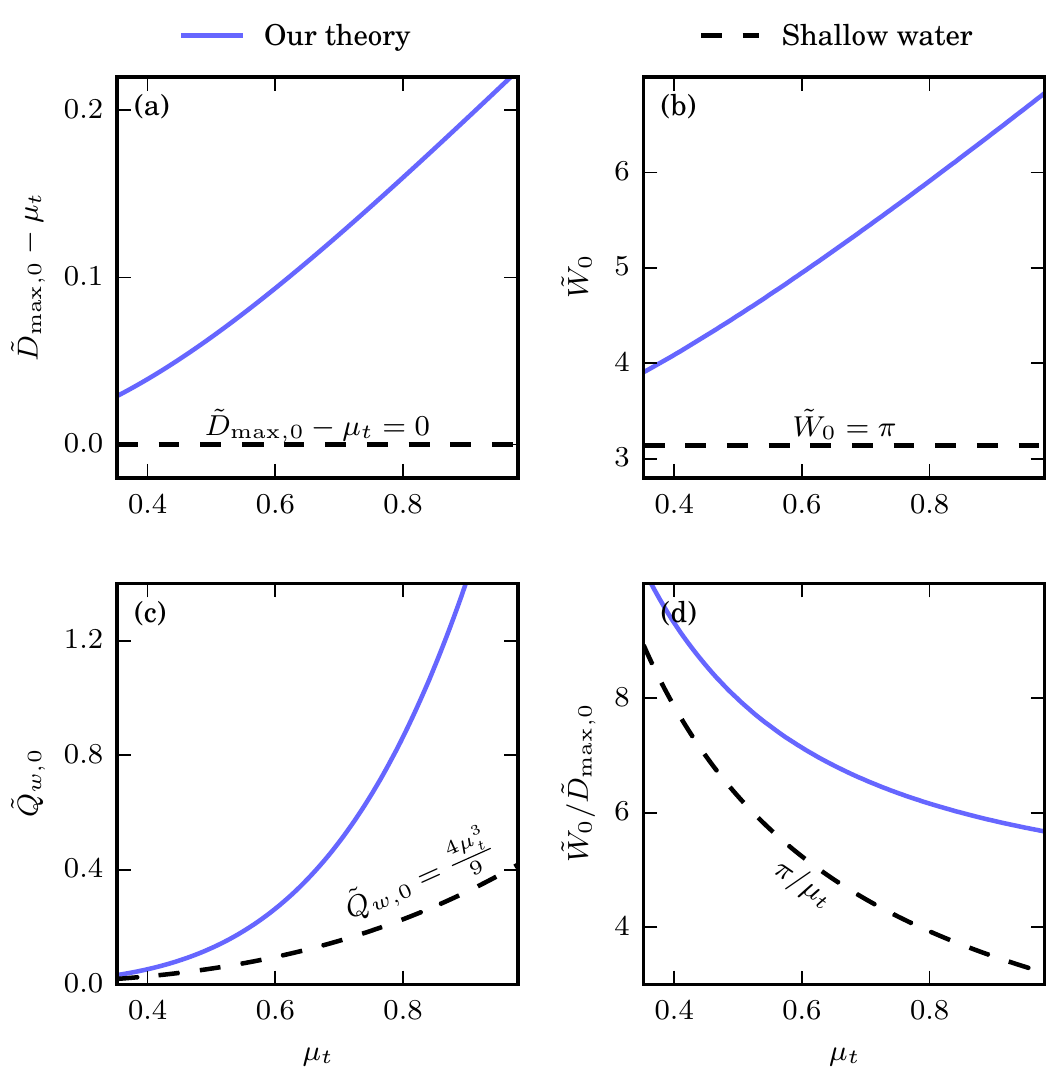}
\caption{Properties of non-dimensional inert rivers as a function of the friction coefficient, $\mu_t$. The blue lines correspond to our model (Eq. \ref{Inert_BVP_SI}), while the black dashed lines correspond to the shallow-water inert river (Eq. \ref{Inert_SW_sol_SI}). (a) Difference between non-dimensional depth and the friction coefficient, $\tilde{D}_{\text{max},0} - \mu_t$. Black dashed line corresponds to $\tilde{D}_{\text{max},0} - \mu_t = 0$. (b) Non-dimensional width, $\tilde{W}_{0}$. Black dashed line corresponds to $\tilde{W}_{0} = \pi$. (c) Non-dimensional fluid discharge, $\tilde{Q}_{w,0} \equiv \int \tilde{D}^3/3 \text{d}\tilde{y}$. Black dashed line corresponds to $\tilde{Q}_{w,0} = 4\mu_t^3/9$. (d) Aspect ratio, $\tilde{W}_{0} / \tilde{D}_{\text{max},0}$. Black dashed line corresponds to $\tilde{W}_0 / \tilde{D}_{\text{max},0} = \pi/\mu_t$.   }
\label{fig:dependence_on_mt}
\end{figure}

In the Parker and weak transport regimes, the shape of the river is determined by that of an inert river. We cannot find the properties of an inert river in our model analytically; instead, to find its profile, $\tilde{D}_0(\tilde{y})$, we have to solve Eq. 17 of the main text with $\tilde{\xi} \rightarrow \infty$:
\begin{linenomath*}
\begin{equation}\label{Inert_BVP_SI}
\sqrt{ \left( \tilde{D}_0 + \tfrac{1}{3} (\tilde{D}_0^3 )'' \right)^2 +  \tilde{D}_0'^2 } - \mu_t =  0  \text{  . }
\end{equation}
\end{linenomath*}
The only parameter in this equation is $\mu_t$, so the inert river shape only depends on this friction coefficient. In Fig. \ref{fig:dependence_on_mt}, we show how its properties --- depth, $\tilde{D}_{\text{max},0}$, width, $\tilde{W}_0$, non-dimensional fluid discharge, $\tilde{Q}_{w,0}$, and aspect ratio, $\tilde{W}_0 / \tilde{D}_{\text{max},0}$ --- depend on $\mu_t$ (blue lines in Fig. \ref{fig:dependence_on_mt}). 

We can understand how these properties generally depend on $\mu_t$ by comparing them to the inert, shallow-water river whose shape can be found analytically (black dashed lines in Fig. \ref{fig:dependence_on_mt}). After neglecting the momentum diffusion, Eq. \ref{Inert_BVP_SI} becomes
\begin{linenomath*}
\begin{equation}\label{Inert_SW_SI}
\sqrt{ \tilde{D}_{\text{sw},0}^2 +  \tilde{D}_{\text{sw},0}'^2 } - \mu_t =  0  \text{  , }
\end{equation}
\end{linenomath*}
where $\tilde{D}_{\text{sw},0}$ is the depth of an inert shallow-water river. This equation has a simple solution \cite{seizilles2013width}
\begin{linenomath*}
\begin{equation}\label{Inert_SW_sol_SI}
\tilde{D}_{\text{sw},0} = \mu_t \cos \tilde{y}  \text{  . }
\end{equation}
\end{linenomath*}
From here, we find that the shallow-water inert river has $\tilde{D}_{\text{max},\text{sw},0} = \mu_t$, $\tilde{W}_{\text{sw},0} = \pi$, $\tilde{Q}_{w,\text{sw},0} = 4\mu_t^3/9$, and an aspect ratio of $\tilde{W}_{\text{sw},0} / \tilde{D}_{\text{max},\text{sw},0} = \pi/\mu_t$. These values represent bounds for our model --- Fig. \ref{fig:dependence_on_mt} shows that the inert river in our model has a depth greater than $\mu_t$, a width greater than $\pi$, a non-dimensional fluid discharge greater than $4\mu_t^3/9$, and an aspect ratio greater than $\pi/\mu_t$. For small $\mu_t$, our model approaches the shallow-water inert river, while, as we increase $\mu_t$, the non-dimensional inert river in our model becomes less and less like the shallow-water one.  

\subsection{Limiting river depth, $\tilde{D}_{\mathrm{max},c}$}\label{sec:limiting_depth_SI}

\begin{figure}
\centering
\includegraphics[width=0.5\textwidth]{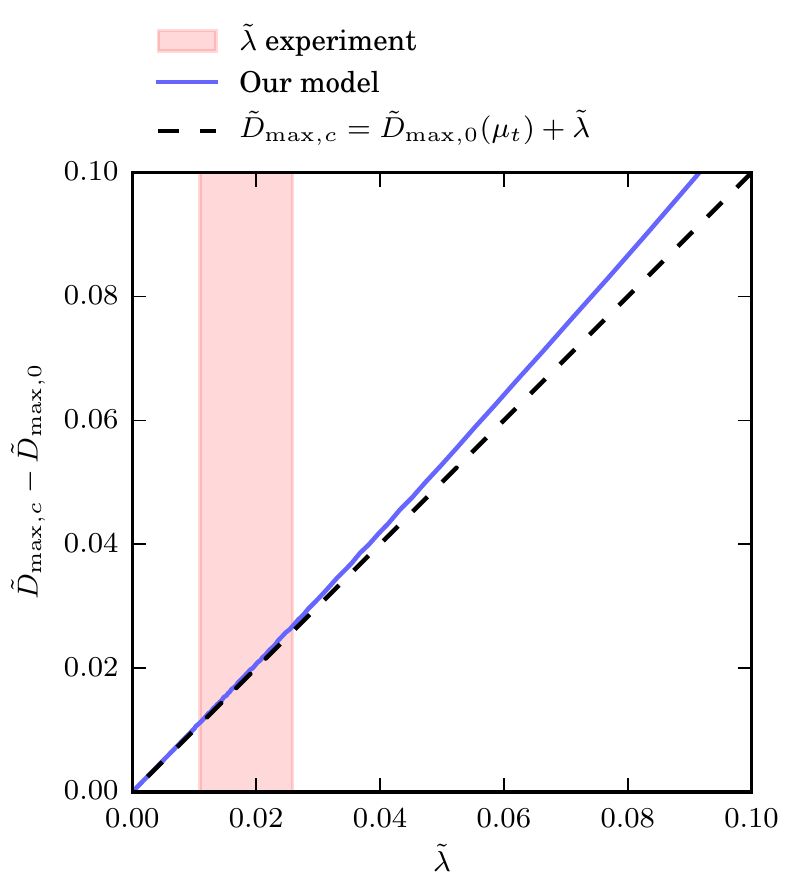}
\caption{  The dependence of limiting river depth, $\tilde{D}_{\text{max},c}$, on $\tilde{\lambda}$. The vertical axis shows the difference between the limiting and inert river depth, $\tilde{D}_{\text{max},c} - \tilde{D}_{\text{max},0}$, while the horizontal axis is the value of $\tilde{\lambda}$ ($\mu_t = 0.9$ is fixed). The blue line results from numerical solution of Eq. 17 of the main text. The black dashed line represents the function $\tilde{D}_{\text{max}, c} - \tilde{D}_{\text{max}, 0} = \tilde{\lambda}$. The red shading represents the estimate and the variability of $\tilde{\lambda}$ observed in the experiments.   }
 \label{fig:plot_delta_D_vs_lambda_1}
\end{figure}

In the main text, we noted that the value of the non-dimensional depth of an infinite, flat river, $\tilde{D}_{\text{max},c}$, depends on parameters $\mu_t$ and $\tilde{\lambda}$ (section \textit{Inert, active, and limiting river} of the main text), and we explained that for vanishing $\tilde{\lambda}$, $\tilde{D}_{\text{max},c}$ equals the inert river depth, $\tilde{D}_{\text{max},0}(\mu_t)$ (section \textit{The Parker regime} of the main text). In Fig. \ref{fig:plot_delta_D_vs_lambda_1}, we show how $\tilde{D}_{\text{max},c}(\tilde{\lambda},\mu_t)$ depends on $\tilde{\lambda}$ with $\mu_t$ fixed. There, we numerically show that, for small $\tilde{\lambda}$, the depth $\tilde{D}_{\text{max},c}$ behaves as
\begin{linenomath*}
\begin{equation}\label{Dc_infty_SI}
\tilde{D}_{\text{max}, c}(\tilde{\lambda},\mu_t) \approx \tilde{D}_{\text{max},0}(\mu_t)  + \tilde{\lambda} \quad \text{ for } \quad \tilde{\lambda} \ll 1 \text{  . }
\end{equation}
\end{linenomath*}
For experimental parameters ($\mu_t = 0.9$ and $\tilde{\lambda} = 0.02$), we find that the relative error between $\tilde{D}_{\text{max},c}$ and Eq. \ref{Dc_infty_SI} is about $0.06\%$. Although we do not understand exactly why this relationship holds, we found it to be true for all the values of $\mu_t$ we tested.

\newpage


\section{Parker regime}

The Parker regime corresponds to a river with no sediment diffusion that splits its channel into inert banks and a flat, active bottom (section \textit{The Parker regime} of the main text). Such a river has a rectangular sediment flux profile of width $W_T^{(P)}$ and height $q_s^{(P)}$.

\begin{figure}
\centering
\includegraphics[width=0.5\textwidth]{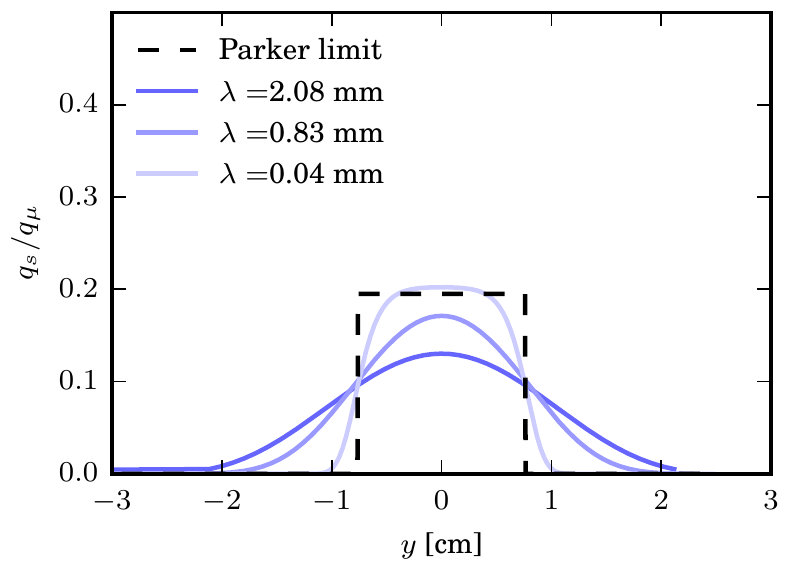}
\caption{Sediment flux profile of a river in our model with $Q_s = 30 \text{ grains s}^{-1}$ and $Q_w = 1 \text{ l min}^{-1}$ and the varying sediment diffusion length scale, $\lambda$. The blue lines are the numerical sediment flux profiles for different values of $\lambda$, while the black dashed line is the Parker regime (Eqs. \ref{Parker_river_main_text_SI}) in the limit $\lambda \rightarrow 0$. }
\label{fig:reaching_parker_limit}
\end{figure}

In Fig. \ref{fig:reaching_parker_limit}, we show that the Parker regime is, in fact, the limit of our model when $\lambda \rightarrow 0$. The sediment flux profile of Eq. 13 of the main text approaches the rectangular shape of the Parker regime as $\lambda$ decreases. This is to be expected --- when the sediment diffusion length scale, $\lambda$, vanishes, the sediment flux, $q_s = q_\mu \exp[(D-\xi)/\lambda]$, also vanishes for $D < \xi$, and becomes infinite for $D > \xi$. Therefore, to have a meaningful solution when $\lambda \rightarrow 0$, the river bottom must be flat with a depth, $D_\text{max} = \xi$. Physically, vanishing sediment diffusion means that gravity pulls each moving grain of sediment towards the river bottom from which it cannot escape by random collisions with the bed. When $\lambda$ is finite, the region over which inert banks transition to the flat bottom always has a finite size of the order $\lambda$. When the water and sediment discharges are large, both the bank width, $W_0$, and the transport width, $W_T$, are large too, so we can neglect the transition region: this is the gist of the Parker regime.

The depth, sediment flux, transport width, and total width of a river in the Parker regime are (as we explained in section \textit{The Parker regime} of the main text):
\begin{linenomath*}
\begin{equation}\label{Parker_river_main_text_SI}
D_\text{max}^{(P)} = \frac{L_s}{S^{(P)}} \tilde{D}_{\text{max},0}  \quad \text{  , } \quad q_s^{(P)} = q_\mu (\tilde{D}_{\text{max},0} - \mu_t) \quad \text{  , } \quad W_T^{(P)} = \frac{Q_s}{q_s^{(P)}} \quad \text{  , } \quad W^{(P)} = \frac{L_s}{S^{(P)}}\tilde{W}_0 +  \frac{Q_s}{q_s^{(P)}} \text{  . }
\end{equation}
\end{linenomath*}

To get the above quantities, we need to find the downstream slope, $S^{(P)}$. This follows from the fluid discharge constraint. According to Eq. 15 of the main text, the fluid discharge is
\begin{linenomath*}
\begin{equation}\label{Qw_Parker_SI}
Q_w = \frac{gS^{(P)}}{\nu}\int_{-W/2}^{W/2} \frac{D^3}{3} \text{d}y = Q_{w,0} + \frac{g S^{(P)}}{\nu} \frac{  D_\text{max}^{(P)3} }{3}W_T^{(P)} \text{  , }
\end{equation}
\end{linenomath*}
where $Q_{w,0} \equiv \int_{-W_0/2}^{W_0/2} \frac{D_0^3}{3} \text{d}y$ is the contribution from the inert banks. A dimensional fluid discharge, $Q_w$, can be related to the non-dimensional one, $\tilde{Q}_w = \int \frac{\tilde{D}^3}{3} \text{d}\tilde{y} $, through
\begin{linenomath*}
\begin{equation}\label{nondim_Qw_SI}
Q_w = \frac{ g L_s^4 }{ \nu S^3} \tilde{Q}_w \text{  , }
\end{equation} 
\end{linenomath*}
so that Eq. \ref{Qw_Parker_SI} becomes
\begin{linenomath*}
\begin{equation}\label{Qw_Parker_SI_2}
Q_w = \frac{ g L_s^4 }{ \nu S^{(P)3}} \tilde{Q}_{w,0} + \frac{g S^{(P)}}{\nu} \frac{  D_\text{max}^{(P)3} }{3}W_T^{(P)} \text{  , }
\end{equation}
\end{linenomath*}
where $\tilde{Q}_{w,0} \equiv \int_{-\tilde{W}_0/2}^{\tilde{W}_0/2} \frac{\tilde{D}_0^3}{3} \text{d}\tilde{y}$ depends only on $\mu_t$. Combining Eq. \ref{Qw_Parker_SI_2} with Eqs. \ref{Parker_river_main_text_SI}, we retrieve Eq. 24 of the main text
\begin{linenomath*}
\begin{equation}\label{Parker_L_SI}
Q_w = \frac{ g L_s^4 }{ \nu S^{(P)3}} \left( \tilde{Q}_{w,0} + \frac{Q_s S^{(P)} \tilde{D}_{\text{max},0}^3}{3 q_\mu  L_s  (\tilde{D}_{\text{max},0} - \mu_t) } \right) \text{  . }
\end{equation}
\end{linenomath*}
Depending on the physical parameters that enter it, Eq. \ref{Parker_L_SI} can have multiple solutions for $S^{(P)}$, but there is always only one positive, real solution.

\newpage

\section{Estimating sediment discharge based on river geometry}

Our model predicts a link between the river geometry and its sediment load. Here, we will use our experimental dataset to show how to estimate the sediment discharge of a river from the shape of its cross-section.

\begin{figure}
\centering
\includegraphics[width=0.5\textwidth]{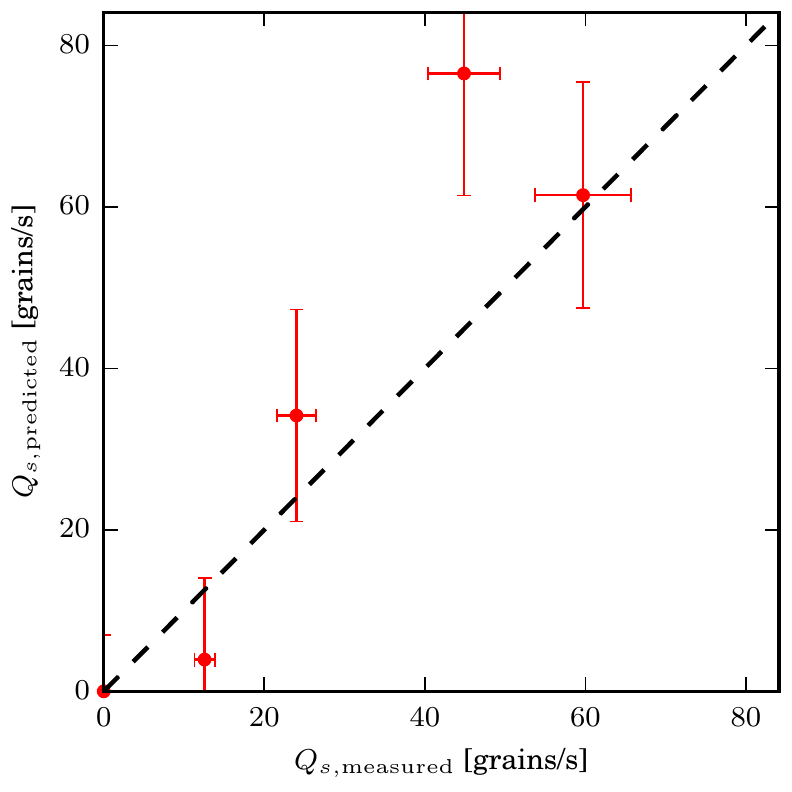}
\caption{Comparing the measured sediment discharge (x-axis) to that estimated using Eq. 22 of the main text (y-axis). Black dashed line is a one-to-one line. }
\label{fig:plot_measured_predicted_Qs_experiments_1}
\end{figure}

Equation 22 of the main text relates the sediment discharge to river depth and width, assuming we know the universal quantities, such as the limiting sediment flux and the inert river aspect ratio ($q_s^{(P)}$ and $\tilde{W}_0/\tilde{D}_{\mathrm{max},0}$ in Eq. 22). Although these universal quantities follow from the theory, various approximations we made may make them inaccurate. Moreover, for a natural river, for which the theory is not yet available, these quantities are unknown a priori. On an ensemble of rivers with different water and sediment discharges, $q_s^{(P)}$ is the maximum allowed sediment flux while $\tilde{W}_0/\tilde{D}_{\mathrm{max},0}$ is the minimum allowed aspect ratio. Therefore, instead of using our theoretical predictions, here we estimate $q_s^{(P)}$ and $\tilde{W}_0/\tilde{D}_{\mathrm{max},0}$ as the maximum sediment flux and minimum aspect ratio from our experimental dataset. Thus, we estimate $q_s^{(P)}|_\mathrm{exp.} \approx 27.4 \text{ grains cm}^{-1}\text{s}^{-1}$ and $\tilde{W}_0/\tilde{D}_{\mathrm{max},0}|_\mathrm{exp.} \approx 4.3$. Using these values, we show in Fig. \ref{fig:plot_measured_predicted_Qs_experiments_1} that the sediment discharge estimated using Eq. 22 of the main text falls within the uncertainty range of the measurements.

\newpage

\section{Effects of momentum diffusion}

In section \textit{The Parker regime} of the main text, we claimed that momentum diffusion is essential for the existence of a steady-state channel with a non-vanishing sediment discharge, as first suggested by \citet{parker1978self}. In this section, we will first illustrate this statement by comparing our model to the shallow-water river without momentum diffusion. Then, we will investigate the effect of momentum diffusion in detail by considering the forces acting on the river bed. 

\subsection{Comparison with shallow-water rivers}

\begin{figure}
\centering
\includegraphics[width=1.\textwidth]{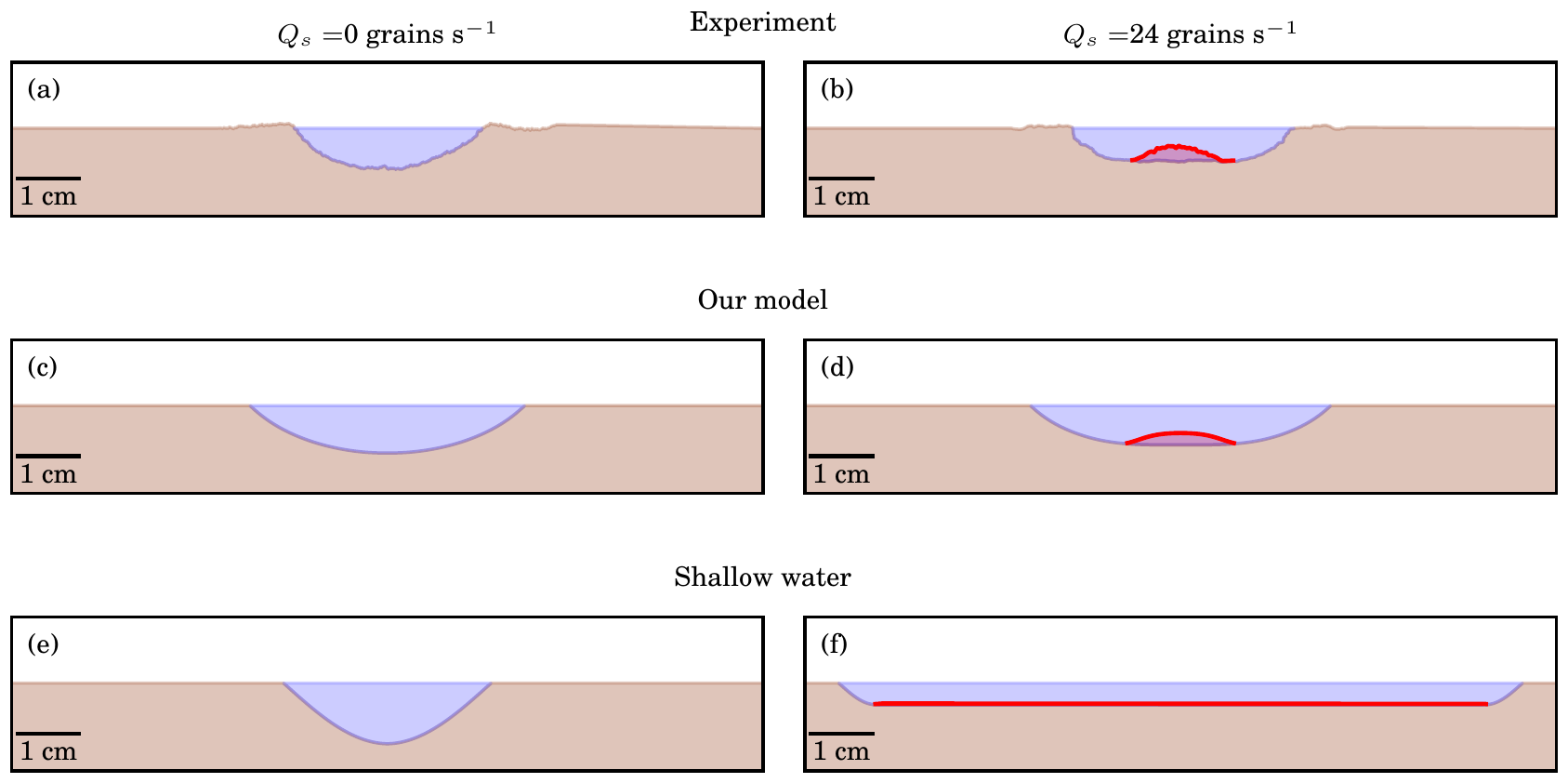}
\caption{River profiles in the experiment (panels (a) and (b)), our model (panels (c) and (d)), and shallow-water approximation (panels (e) and (f)). Profiles in the left column (panels (a), (c), and (e)) are inert rivers, while profiles in the right column (panels (b), (d), and (f)) are active rivers with $Q_s = 24 \text{ grains s}^{-1}$. Red lines in panels (b), (d), and (f) represent the sediment flux profile where the zero, $q_s = 0$, is vertically shifted in the plots to coincide with the river bottom. While inert rivers in both our model and the shallow-water model are comparable with experiments, momentum diffusion is necessary to capture the shape of active rivers. }
\label{fig:plot_river_profiles_with_SW}
\end{figure}

The shallow-water model, which neglects the cross-stream diffusion of momentum entirely, is defined by a first order ordinary differential equation:
\begin{linenomath*}
\begin{equation}\label{Active_SW_SI}
\sqrt{ \tilde{D}_\text{sw}^2 +  \tilde{D}_\text{sw}'^2 } - \mu_t =  e^{( \tilde{D}_\text{sw} - \tilde{\xi} ) / \tilde{\lambda} }  \text{  . }
\end{equation}
\end{linenomath*}
When $\tilde{\xi} \rightarrow \infty$, the right-hand side vanishes and the model reduces to Eq. \ref{Inert_SW_SI} for an inert shallow-water river, which has been successfully used to predict the shape of experimental inert rivers \cite{seizilles2013width}. We compare the inert shallow-water river with the experiment and our model in Figs. \ref{fig:plot_river_profiles_with_SW}a, c, and e. There, we can see that, while there exists some difference between the models and the experiment, they are largely comparable to each other.

However, when sediment transport is included ($\tilde{\xi} < \infty$), this model is insufficient to reproduce the experiments. We compare an active experimental river with our model and the shallow-water model in Figs. \ref{fig:plot_river_profiles_with_SW}b, d, and f. We can see that, while our model and the experiment seem comparable to each other, the shallow-water model is nothing alike. This result may seem somewhat paradoxical --- although we expect that the shallow-water approximation should work better for a wider river, we find that the wider the river, the worse the shallow-water approximation is. We resolve this apparent paradox in the next section.


\subsection{Momentum diffusion in the Parker regime}


In this section, we look closely at the components of the force ratio, $\mu$, to understand the role of momentum diffusion. We explain that momentum diffusion controls the sediment transport through a matching condition at the interface of the river banks and the flat bottom --- a condition that does not depend on the width of the river.

\begin{figure}
\centering
\includegraphics[width=0.5\textwidth]{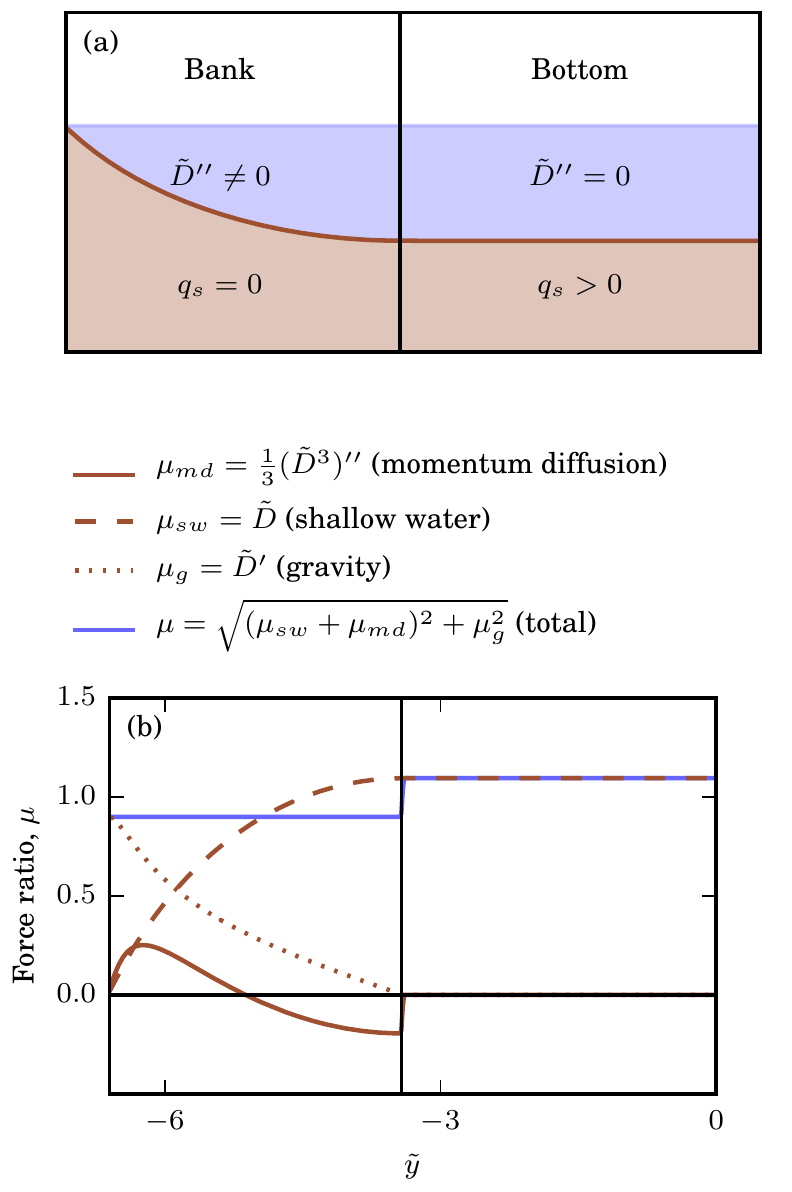}
\caption{Explanation of the mechanism of sediment transport by momentum diffusion. (a) Half a river in our model for $\lambda \rightarrow 0$ and $\mu_t = 0.9$. The river sharply splits into curved, inert banks and a flat, active bottom. (b) Components of the force acting on the bed as a function of the position along the bed. The blue line represents the total force ratio, $\mu = \sqrt{(\mu_\text{sw} + \mu_\text{md})^2 + \mu_\text{g}^2}$. The dashed brown line represents the shallow-water contribution to the stress, $\mu_\text{sw} = \tilde{D}$. The solid brown line represents the momentum diffusion contribution to the stress, $\mu_\text{md} = (\tilde{D}^3)''/3$. The dotted brown line represents the gravity contribution, $\mu_\text{g} = \tilde{D}'$. On the banks, all three contributions are non-vanishing and keep the banks at the threshold, $\mu = \mu_t$. On the bottom, only the shallow-water contribution exists and the force ratio is above threshold, $\mu > \mu_t$. The shallow-water and gravity contributions transition continuously from the banks to the bed, but the momentum diffusion contribution experiences a jump which corresponds to a jump in the force ratio, $\mu$. This jump drives sediment transport.}
\label{fig:mechanism_explanation}
\end{figure}

Figure \ref{fig:mechanism_explanation}a shows a river in the Parker regime (in non-dimensional coordinates, for convenience) --- it is sharply split into curved, inert banks, and a flat, active bottom. Figure \ref{fig:mechanism_explanation}b shows how the various components of the force ratio, $\mu$, depend on the position within this river. These components are
\begin{linenomath*}
\begin{align}\label{banks_mu_SI}
&\mu_\text{sw} \equiv \tilde{D}  \text{  ,  } \\
&\mu_\text{md} \equiv \frac{1}{3}(\tilde{D}^3)''  \text{  ,  }\\
&\mu_\text{g} \equiv \tilde{D}' \text{  , } \\
&\mu^2 =  (\mu_\text{sw} + \mu_\text{md})^2 + \mu_\text{g}^2  \text{  , } 
\end{align}
\end{linenomath*}
where $\mu_\text{sw}$ is the shallow-water contribution, $\mu_\text{md}$ is the momentum diffusion contribution, and $\mu_\text{g}$ is the gravity contribution to the total force ratio. The banks are kept at threshold by the combined action of the shallow-water stress, momentum diffusion, and gravity, while the bottom only feels the shallow-water stress. At the point where banks and flat bottom connect the banks have a depth $\tilde{D} = \tilde{D}_\text{max}$, a slope $\tilde{D}' = 0$, and a second derivative $\tilde{D}'' =-\tilde{\kappa} < 0$, where $\tilde{\kappa}$ is the curvature of the banks at their deepest point. The components of the force ratio are, correspondingly, $\mu_\text{sw} = \tilde{D}_\text{max}$, $\mu_\text{g} = 0$, $\mu_\text{md} = -\tilde{D}_\text{max}^2 \tilde{\kappa} < 0$. The threshold condition for the banks is, thus, $\mu_\text{sw} + \mu_\text{md} = \mu_t$, which means that the depth must be greater than the friction coefficient, $\tilde{D}_\text{max} = \mu_\text{sw} = \mu_t - \mu_\text{md} > \mu_t$. The bottom feels only the shallow-water stress, and, therefore, must be above the threshold, $\mu\vert_\text{bottom} = \mu_\text{sw} = \tilde{D}_\text{max} > \mu_t$. 


Therefore, the absence of sediment transport on the banks translates into excess stress on the flat bottom. The sediment flux on the bottom, $q_{s} / q_\mu = \mu \vert_\text{bottom} - \mu_t =  -  \mu_\text{md}$, corresponds to the jump in the momentum diffusion contribution due to the sudden disappearance of curvature. Thus, the sediment flux is driven by the boundary condition at the junction of the curved banks with the flat bottom. 

Written in terms of depth and its derivatives, the sediment flux is $q_{s} / q_\mu =  \tilde{D}_\text{max}^2 \tilde{\kappa}$. The curvature, $\tilde{\kappa} \equiv -\tilde{D}''$, scales roughly as $\tilde{\kappa} \sim \tilde{D}_{\text{max},0} / \tilde{W}_0^2$, so the sediment flux scales as $q_{s} / q_\mu \sim \tilde{D}_{\text{max},0}^3 / \tilde{W}_0^2$. Taking the depth, $\tilde{D}_{\text{max},0}$, to be of order one, the sediment flux is inversely proportional to the square of the aspect ratio of an inert river, $W_0^2 / D_{\text{max},0}^2$. \citet{parker1978self} likewise found that the distance to threshold is inversely proportional to the aspect ratio squared --- a signature of the cross-stream diffusion of momentum.  

Without cross-stream momentum diffusion, there can be no sediment transport in the limit $\lambda \rightarrow 0$ in our model. When $\lambda$ is small but non-vanishing, the maximal sediment flux is $q_{s,c} / q_\mu \approx \tilde{D}_{\text{max}, 0}(\mu_t)  - \mu_t + \tilde{\lambda}$, where we used Eq. \ref{Dc_infty_SI} for the depth of the infinite, limiting river, $\tilde{D}_{\text{max}, c}$. The first term in this equation, $\tilde{D}_{\text{max}, 0}(\mu_t) - \mu_t$, is the contribution to sediment transport from the momentum diffusion, while the second term, $\tilde{\lambda}$, is the contribution from the sediment diffusion. For experimental parameters ($\mu_t = 0.9$ and $\tilde{\lambda} = 0.02$), we thus find that momentum diffusion is responsible for about 90\% of the sediment transport.

\newpage

\bibliography{bibRivers}